\documentclass[sigconf,screen=true]{acmart}

\AtBeginDocument{%
  \providecommand\BibTeX{{%
    \normalfont B\kern-0.5em{\scshape i\kern-0.25em b}\kern-0.8em\TeX}}}

\setcopyright{acmcopyright}
\copyrightyear{2021}
\acmYear{2021}
\acmDOI{TBA}

\copyrightyear{2021}
\acmYear{2021}
\setcopyright{acmlicensed}\acmConference[IMC '21]{ACM Internet Measurement Conference}{November 2--4, 2021}{Virtual Event, USA}
\acmBooktitle{ACM Internet Measurement Conference (IMC '21), November 2--4, 2021, Virtual Event, USA}
\acmPrice{15.00}
\acmDOI{10.1145/3487552.3487811}
\acmISBN{978-1-4503-9129-0/21/11}

\usepackage{amsmath}
\usepackage{xspace}
\usepackage{booktabs}
\usepackage{url}
\usepackage{amsfonts}
\usepackage{multirow}
\usepackage[ruled,vlined]{algorithm2e}
\usepackage{enumitem}
\usepackage{numprint}
\usepackage{tcolorbox}
\usepackage{xurl}
\usepackage{xstring}

\usepackage{xcolor}
\usepackage{subcaption}

\usepackage[english]{babel}
\newcommand{\etal}{\textit{et al.}}
\npthousandsep{,}
\npdecimalsign{.}

\newcommand{\empirical}[1]{#1}

\newcommand{\block}[1]{\href{https://etherscan.io/block/#1}{#1}}
\newcommand{\account}[1]{\href{https://etherscan.io/address/#1}{\wrapletters{#1}}}
\newcommand{\etherscantx}[1]{\href{https://etherscan.io/tx/#1}{\wrapletters{#1}}}

\newcommand*\wrapletters[1]{\wr@pletters#1\@nil}
\def\wr@pletters#1#2\@nil{#1\allowbreak\if&#2&\else\wr@pletters#2\@nil\fi}

\newcommand{\point}[1]{\par\smallskip\noindent\textbf{#1:}\xspace}

\usepackage{listings}

\definecolor{lightgray}{rgb}{.9,.9,.9}
\definecolor{darkgray}{rgb}{.4,.4,.4}
\definecolor{purple}{rgb}{0.65, 0.12, 0.82}

\lstdefinelanguage{JavaScript}{
  keywords={typeof, new, true, false, catch, function, return, null, catch, switch, var, if, in, while, do, else, case, break},
  keywordstyle=\color{blue}\bfseries,
  ndkeywords={class, export, boolean, throw, implements, import, this},
  ndkeywordstyle=\color{darkgray}\bfseries,
  identifierstyle=\color{black},
  sensitive=false,
  comment=[l]{//},
  morecomment=[s]{/*}{*/},
  commentstyle=\color{purple}\ttfamily,
  stringstyle=\color{red}\ttfamily,
  morestring=[b]',
  morestring=[b]"
}

\definecolor{verylightgray}{rgb}{.97,.97,.97}

\lstdefinelanguage{Solidity}{
	keywords=[1]{anonymous, assembly, assert, balance, break, call, callcode, case, catch, class, constant, continue, constructor, contract, debugger, default, delegatecall, delete, do, else, emit, event, experimental, export, external, false, finally, for, function, gas, if, implements, import, in, indexed, instanceof, interface, internal, is, length, library, log0, log1, log2, log3, log4, memory, modifier, new, payable, pragma, private, protected, public, pure, push, require, return, returns, revert, selfdestruct, send, solidity, storage, struct, suicide, super, switch, then, this, throw, transfer, true, try, typeof, using, value, view, while, with, addmod, ecrecover, keccak256, mulmod, ripemd160, sha256, sha3}, 
	keywordstyle=[1]\color{blue}\bfseries,
	keywords=[2]{address, bool, byte, bytes, bytes1, bytes2, bytes3, bytes4, bytes5, bytes6, bytes7, bytes8, bytes9, bytes10, bytes11, bytes12, bytes13, bytes14, bytes15, bytes16, bytes17, bytes18, bytes19, bytes20, bytes21, bytes22, bytes23, bytes24, bytes25, bytes26, bytes27, bytes28, bytes29, bytes30, bytes31, bytes32, enum, int, int8, int16, int24, int32, int40, int48, int56, int64, int72, int80, int88, int96, int104, int112, int120, int128, int136, int144, int152, int160, int168, int176, int184, int192, int200, int208, int216, int224, int232, int240, int248, int256, mapping, string, uint, uint8, uint16, uint24, uint32, uint40, uint48, uint56, uint64, uint72, uint80, uint88, uint96, uint104, uint112, uint120, uint128, uint136, uint144, uint152, uint160, uint168, uint176, uint184, uint192, uint200, uint208, uint216, uint224, uint232, uint240, uint248, uint256, var, void, ether, finney, szabo, wei, days, hours, minutes, seconds, weeks, years},	
	keywordstyle=[2]\color{teal}\bfseries,
	keywords=[3]{block, blockhash, coinbase, difficulty, gaslimit, number, timestamp, msg, data, gas, sender, sig, value, now, tx, gasprice, origin},	
	keywordstyle=[3]\color{violet}\bfseries,
	identifierstyle=\color{black},
	sensitive=false,
	comment=[l]{//},
	morecomment=[s]{/*}{*/},
	commentstyle=\color{gray}\ttfamily,
	stringstyle=\color{red}\ttfamily,
	morestring=[b]',
	morestring=[b]"
}

\newcommand{\CrawlingEndBlock}{\empirical{\block{12344944}}\xspace}
\newcommand{\AccumulativeLiquidatedAmount}{\empirical{$\numprint{807.46}$M}~USD\xspace}

\newcommand{\TotalSuccessfulLiquidations}{\empirical{$\numprint{28138}$}\xspace}
\newcommand{\AaveLiquidations}{\empirical{$\numprint{3809}$}\xspace}
\newcommand{\AaveVTLiquidations}{\empirical{$\numprint{1039}$}\xspace}
\newcommand{\CompoundLiquidations}{\empirical{$\numprint{6766}$}\xspace}
\newcommand{\dYdXLiquidations}{\empirical{$\numprint{9762}$}\xspace}
\newcommand{\MakerDAOAuctions}{\empirical{$\numprint{6762}$}\xspace}

\newcommand{\TotalNumLiquidators}{\empirical{$\numprint{2011}$}\xspace}
\newcommand{\AaveNumLiquidators}{\empirical{$\numprint{665}$}\xspace}
\newcommand{\AaveVTNumLiquidators}{\empirical{$\numprint{125}$}\xspace}
\newcommand{\CompoundNumLiquidators}{\empirical{$\numprint{657}$}\xspace}
\newcommand{\dYdXNumLiquidators}{\empirical{$\numprint{600}$}\xspace}
\newcommand{\MakerDAONumLiquidators}{\empirical{$\numprint{140}$}\xspace}

\newcommand{\TotalLiquidationProfit}{\empirical{$63.59$M}~USD\xspace}

\newcommand{\TotalPerLiquidatorProfit}{\empirical{$\numprint{31.62}$K}~USD\xspace}
\newcommand{\AavePerLiquidatorProfit}{\empirical{$\numprint{10.76}$K}~USD\xspace}
\newcommand{\AaveVTPerLiquidatorProfit}{\empirical{$\numprint{43.12}$K}~USD\xspace}
\newcommand{\CompoundPerLiquidatorProfit}{\empirical{$\numprint{39.94}$K}~USD\xspace}
\newcommand{\dYdXPerLiquidatorProfit}{\empirical{$\numprint{14.30}$K}~USD\xspace}
\newcommand{\MakerDAOPerLiquidatorProfit}{\empirical{$\numprint{115.84}$K}~USD\xspace}

\newcommand{\BiggestSingleLiquidationProfit}{\empirical{$\numprint{4.04}$M}~USD\xspace}

\newcommand{\MostActiveLiquidatorLiquidations}{\empirical{$\numprint{2482}$}\xspace}
\newcommand{\MostActiveLiquidatorProfit}{\empirical{$\numprint{741.75}$K}~USD\xspace}
\newcommand{\MostLucrativeLiquidatorLiquidations}{\empirical{$\numprint{112}$}\xspace}
\newcommand{\MostLucrativeLiquidatorProfit}{\empirical{$\numprint{5.84}$M}~USD\xspace}

\newcommand{\NonProfitableMakerDAOLiquidations}{\empirical{$\numprint{641}$}\xspace}
\newcommand{\NonProfitableMakerDAOLoss}{\empirical{$\numprint{467.44}$K}~USD\xspace}

\newcommand{\MakerDAOAverageAuctionDuration}{\empirical{$\numprint{2.06}\pm\numprint{6.43}$}~hours\xspace}
\newcommand{\MakerDAOAuctionsInOneHour}{\empirical{$\numprint{4173}$}\xspace}
\newcommand{\LongestMakerDAOAuction}{\empirical{$\numprint{346.67}$}~hours\xspace}
\newcommand{\LongestMakerDAOAuctionLastBid}{\empirical{$\numprint{344.60}$}~hours\xspace}
\newcommand{\MakerDAOAuctionsWithMoreThanOneBids}{\empirical{$\numprint{4537}$}\xspace}
\newcommand{\MakerDAOFirstBidInterval}{\empirical{$\numprint{4.12}\pm\numprint{25.52}$}~minutes\xspace}
\newcommand{\MakerDAOBidInterval}{\empirical{$\numprint{38.97}\pm\numprint{89.34}$}~minutes\xspace}
\newcommand{\MakerDAOTerminateInTend}{\empirical{$\numprint{3377}$}\xspace}
\newcommand{\MakerDAOTerminateInDent}{\empirical{$\numprint{3385}$}\xspace}
\newcommand{\MakerDAOAverageBidders}{\empirical{$\numprint{1.99}$}\xspace}
\newcommand{\MakerDAOBidsPerAuction}{\empirical{$\numprint{2.63}\pm\numprint{1.96}$}\xspace}
\newcommand{\MakerDAOTendBidsPerAuction}{\empirical{$\numprint{1.58}\pm\numprint{0.95}$}\xspace}
\newcommand{\MakerDAODentBidsPerAuction}{\empirical{$\numprint{1.06}\pm\numprint{1.62}$}\xspace}

\newcommand{\NumLiquidationsHigherThanAverageGasPrice}{\empirical{$\numprint{73.97}\%$}\xspace}

\newcommand{\TotalFlashLoans}{\empirical{$\numprint{623}$}\xspace}
\newcommand{\AavedYdXFlashLoans}{\empirical{$\numprint{320}$}\xspace}
\newcommand{\AaveVTAaveVTFlashLoans}{\empirical{$\numprint{61}$}\xspace}
\newcommand{\AaveVTdYdXFlashLoans}{\empirical{$\numprint{97}$}\xspace}
\newcommand{\CompoundAaveFlashLoans}{\empirical{$\numprint{114}$}\xspace}
\newcommand{\CompounddYdXFlashLoans}{\empirical{$\numprint{31}$}\xspace}

\newcommand{\TotalFlashLoanAmount}{\empirical{$\numprint{483.83}$}M~USD\xspace}
\newcommand{\AavedYdXFlashLoanAmount}{\empirical{$\numprint{93.57}$M}~USD\xspace}
\newcommand{\AaveVTAaveVTlashLoanAmount}{\empirical{$\numprint{1.27}$M}~USD\xspace}
\newcommand{\AaveVTdYdXlashLoanAmount}{\empirical{$\numprint{317.29}$M}~USD\xspace}
\newcommand{\CompoundAaveFlashLoanAmount}{\empirical{$\numprint{32.49}$K}~USD\xspace}
\newcommand{\CompounddYdXFlashLoanAmount}{\empirical{$\numprint{71.67}$M}~USD\xspace}

\newcommand{\HorizontalLiquidations}{\empirical{$\numprint{789}$}\xspace}
\newcommand{\RiseLiquidations}{\empirical{$\numprint{6142}$}\xspace}
\newcommand{\FallLiquidations}{\empirical{$\numprint{5365}$}\xspace}
\newcommand{\RiseFallLiquidations}{\empirical{$\numprint{2142}$}\xspace}
\newcommand{\FallRiseLiquidations}{\empirical{$\numprint{4513}$}\xspace}
\newcommand{\RiseFluctuationLiquidations}{\empirical{$\numprint{4582}$}\xspace}
\newcommand{\FallFluctuationLiquidations}{\empirical{$\numprint{4605}$}\xspace}

\newcommand{\RisePeak}{\empirical{$\numprint{7.33}\%\pm7.26\%$}\xspace}
\newcommand{\FallValley}{\empirical{$-6.70\%\pm6.50\%$}\xspace}
\newcommand{\RiseFallPeak}{\empirical{$1.46\%\pm1.38\%$}\xspace}
\newcommand{\RiseFallValley}{\empirical{$-8.35\%\pm7.90\%$}\xspace}
\newcommand{\FallRisePeak}{\empirical{$5.70\%\pm5.78\%$}\xspace}
\newcommand{\FallRiseValley}{\empirical{$-3.14\%\pm4.70\%$}\xspace}
\newcommand{\RiseFluctuationPeak}{\empirical{$7.85\%\pm7.80\%$}\xspace}
\newcommand{\RiseFluctuationValley}{\empirical{$-4.52\%\pm4.73\%$}\xspace}
\newcommand{\FallFluctuationPeak}{\empirical{$3.49\%\pm4.66\%$}\xspace}
\newcommand{\FallFluctuationValley}{\empirical{$-5.07\%\pm5.86\%$}\xspace}

\newcommand{\OptimalLiquidationAdditionalProfit}{\empirical{$53.96$K}~USD\xspace}
\newcommand{\OptimalLiquidationAdditionalPercentage}{\empirical{$1.36\%$}\xspace}

\newcommand{\MakerDAOMarchProfit}{\empirical{$13.13$M}~USD\xspace}

\newcommand{\ComopoundNovemberProfit}{\empirical{$8.38$M}~USD\xspace}

\newcommand{\ComopoundFebruaryProfit}{\empirical{$9.61$M}~USD\xspace}

\begin{document}

\title[An Empirical Study of DeFi Liquidations]{An Empirical Study of DeFi Liquidations:\\ Incentives, Risks, and Instabilities}

\author{Kaihua Qin}
\email{kaihua.qin@imperial.ac.uk}
\affiliation{%
  \institution{Imperial College London}
  \country{United Kingdom}
}

\author{Liyi Zhou}
\email{liyi.zhou@imperial.ac.uk}
\affiliation{%
  \institution{Imperial College London}
  \country{United Kingdom}
}

\author{Pablo Gamito}
\email{pablo.gamito17@imperial.ac.uk}
\affiliation{%
  \institution{Imperial College London}
  \country{United Kingdom}
}

\author{Philipp Jovanovic}
\email{p.jovanovic@ucl.ac.uk}
\affiliation{%
  \institution{University College London}
  \country{United Kingdom}
}

\author{Arthur Gervais}
\email{a.gervais@imperial.ac.uk}
\affiliation{%
  \institution{Imperial College London}
  \country{United Kingdom}
}

\renewcommand{\shortauthors}{Qin et al.}

\begin{abstract}
Financial speculators often seek to increase their potential gains with leverage. Debt is a popular form of leverage, and with over $39.88$B USD of total value locked (TVL), the Decentralized Finance (DeFi) lending markets are thriving. Debts, however, entail the risks of liquidation, the process of selling the debt collateral at a discount to liquidators. Nevertheless, few quantitative insights are known about the existing liquidation mechanisms.

In this paper, to the best of our knowledge, we are the first to study the breadth of the borrowing and lending markets of the Ethereum DeFi ecosystem. We focus on Aave, Compound, MakerDAO, and dYdX, which collectively represent over~$85$\% of the lending market on Ethereum. Given extensive liquidation data measurements and insights, we systematize the prevalent liquidation mechanisms and are the first to provide a methodology to compare them objectively. We find that the existing liquidation designs well incentivize liquidators but sell excessive amounts of discounted collateral at the borrowers' expenses. We measure various risks that liquidation participants are exposed to and quantify the instabilities of existing lending protocols. Moreover, we propose an optimal strategy that allows liquidators to increase their liquidation profit, which may aggravate the loss of borrowers.
\end{abstract}

\ccsdesc[500]{General and reference~Empirical studies}
\ccsdesc[500]{General and reference~Measurement}
\ccsdesc[300]{Applied computing~Economics}

\keywords{blockchain, DeFi, liquidation}


\maketitle

\section{Introduction}
Cryptocurrencies are notoriously known to attract financial speculators who often seek to multiply their potential monetary upside and financial gains through leverage. Leverage is realized by borrowing assets to perform trades~---~commonly referred to as margin trading. It is apparent that margin trading, speculating with borrowed assets in general, is an incredibly risky endeavor. Yet, the borrowing and lending markets on blockchains are thriving and have reached a collective~$39.88$B USD of {\em total value locked (TVL)} at the time of writing\footnote{\url{https://defipulse.com/}}.

Loans on a blockchain typically operate as follows. Lenders with a surplus of money provide assets to a lending smart contract. Borrowers then provide a security deposit, known as collateral, to borrow cryptocurrency. Because the lending and borrowing on blockchains lacks compulsory means on defaults,
the amount of debt borrowers can take on is typically inferior to the security deposit in value~---~resulting in \emph{over-collateralized loans}. Over-collateralized loans are interesting from a financial perspective, as they enable borrowers to take on leverage.

If the collateral value decreases under a specific threshold (e.g., below $150$\% of the debt value~\cite{makerdao}), the associated debt can be recovered through three means: \emph{(1)} a loan can be made available for liquidation by the smart contract. Liquidators then pay back the debt in exchange for receiving the collateral at a discount (i.e., \emph{liquidation spread}), or the collateral is liquidated through an auction. \emph{(2)} Debt can also be rescued by ``topping up'' the collateral, such that the loan is sufficiently collateralized. \emph{(3)} Finally, the borrower can repay parts of their debt. While users can repay their debts manually, this appears impractical for the average user, as it requires infrastructure to constantly monitor the blockchain, collateral price, and transaction fee fluctuations. For example, even professional liquidation bots from MakerDAO failed to monitor and act upon price variations during blockchain congestion~\cite{maker-fail}. 

In this paper we make the following contributions. 

\begin{enumerate}
    \item {\bf Liquidation Models and Insights:} We provide the first longitudinal study of the four major lending platforms MakerDAO, Aave, Compound, and dYdX, capturing collectively over $85$\% of the borrowing/lending market on the Ethereum blockchain. By focusing on the protocol's liquidation mechanisms, we systematize their liquidation designs. MakerDAO, for instance, follows an auction-based liquidation process, while Aave, Compound, and dYdX operate under a fixed spread liquidation model.
    
    \item {\bf Data Analytics:} We provide on-chain data analytics covering the entire existence of the four protocols ($2$ years). Our findings show how the accumulative liquidation proceeds amount to~\AccumulativeLiquidatedAmount, we identify~\TotalNumLiquidators unique liquidator addresses and~\TotalSuccessfulLiquidations liquidation events, of which~\NonProfitableMakerDAOLiquidations auction liquidations are not profitable for the liquidators. We show how~\NumLiquidationsHigherThanAverageGasPrice of the liquidations pay an above average transaction fee, indicating competitive behavior. We find the existence of bad debts, the borrowing positions that do not financially incentivize borrowers to perform a close. Notably, Aave V2 has accumulated up to~$87.4$K~USD of bad debts by the end of April,~2021.
    We quantify how sensitive debt behaves to collateral price declines and find that, for example, a~$43$\% reduction of the ETH price (analogous to the ETH price decline on the~13th of March,~2020) would engender liquidatable collateral volume of~$1.07$B USD on MakerDAO.
    \item {\bf Objective Liquidation Mechanism Comparison:} We provide a methodology to compare quantitatively whether a liquidation mechanism favors a borrower or a liquidator. We find evidence that fixed spread liquidation mechanisms favor liquidators over borrowers. That is, because existing DeFi systems are parameterized to allow more collateral than necessary to be liquidated.
    
    \item {\bf Optimal Fixed Spread Liquidation Strategy:} We propose an optimal fixed spread liquidation strategy. This strategy allows liquidators to lift the restrictions of the close factor (the upper limit of repaid debts in a single liquidation, cf.\ Section~\ref{sec:terminology}) within two successive liquidations. We provide a case study of a past liquidation transaction and show that the optimal strategy could have increased the liquidation profit by \OptimalLiquidationAdditionalProfit (\OptimalLiquidationAdditionalPercentage), validated through concrete execution on the real blockchain state. This optimal strategy can further aggravate the loss of borrowers.
\end{enumerate}

The remainder of the paper is organized as follows. Section~\ref{sec:background} outlines the background on blockchain and lending, while we systematize existing liquidation mechanisms in Section~\ref{sec:existing-protocols}. Section~\ref{sec:insights} provides liquidation data insights from empirical data. 
We discuss how to objectively compare liquidation mechanisms and the optimal liquidation strategy in Section~\ref{sec:better-liquidation}. 
We outline related work in Section~\ref{sec:related-work} and conclude the paper in Section~\ref{sec:conclusion}.

\section{Lending on the Blockchain}\label{sec:background}
We proceed by outlining the required background on blockchain and DeFi for the remainder of the paper.

\subsection{Blockchain \& Smart Contract}
Blockchains are distributed ledgers that enable peers to transact without the need to entrust third-party intermediaries. There exist two categories of blockchains: \textit{(i)} permissionless blockchains, where any entity is able to join and leave without permission; \textit{(ii)} permissioned blockchains, which are typically composed of a group of authenticated participants. In this work, we only focus on permissionless blockchains, on top of which DeFi is built.

At its core, a blockchain is a hash-linked chain of blocks operating over a peer-to-peer (P2P) network~\cite{bonneau2015sok}. A block is a timestamped data structure aggregating transactions, which record, e.g., asset transfers. To transfer assets, users need to broadcast digitally signed transactions through the P2P network. The so-called miners then collect transactions, pack transactions into blocks, and append blocks to the blockchain. The whole network follows a consensus protocol (e.g., Nakamoto consensus~\cite{bitcoin}) allowing honest participants to agree on a consistent version of the blockchain. Transactions waiting to be confirmed on-chain are stored in the so-called mempool\footnote{Note that there is no universal mempool across all network participants. Every node maintains its own mempool depending on the received transactions.}. We refer the reader to~\cite{bonneau2015sok,bano2019sok} for a more thorough background on blockchains.

Some blockchains, for example, Ethereum~\cite{wood2014ethereum}, offer generic computation capabilities through smart contracts.
In essence, an Ethereum smart contract is an account controlled by an immutable program (i.e., bytecode). One can trigger the execution of the bytecode by sending a transaction, which contains the executing parameters specified by the transaction sender, to the smart contract account. The EVM, a quasi Turing-complete state machine~\cite{atzei2017survey}, provides the runtime environment to the contract execution. Solidity~\cite{dannen2017introducing}, which can be compiled into bytecode, is to date the most prevalent high-level language for implementing Ethereum smart contracts. Smart contracts are widely used to create cryptocurrencies (also known as tokens) on Ethereum in addition to the native coin ETH. Notably, WETH is a one-to-one equivalent token of ETH.

To submit a transaction on-chain, a user is required to pay a transaction fee. On Ethereum, the transaction fee equals the product of the gas (i.e., an integer measuring the computation complexity of a transaction) and the gas price (i.e., the amount of ETH that the transaction sender is willing to pay for a single unit of gas). Due to the limited space of an Ethereum block (i.e., the total amount of gas consumed in on block), a financially rational miner may include the transactions with the highest gas prices from the mempool into the next block. The blockchain network congests when the mempool grows faster than the transaction inclusion speed due to, for example, traders place substantial orders in a market collapse. Under such circumstances, users have to increase gas prices or wait longer than average to confirm their transactions.

\subsection{Decentralized Finance (DeFi)}
Smart contracts allow, not only the creation of tokens, but further the construction of sophisticated on-chain financial systems, namely Decentralized Finance (DeFi). In DeFi, any entity can design a financial protocol, implement in smart contracts, and deploy on-chain. Compared to traditional finance, DeFi presents promising peculiarities, e.g., non-custody and public verifiability~\cite{qin2021cefi}. Although most DeFi protocols are mirrored services from traditional finance (e.g., exchanges), a proper redesign appears to be necessary considering the special settlement mechanisms of the underlying blockchains. For instance, due to the limited computation capacity, a limit order book with a matching engine, which has been adopted in centralized exchanges for decades, is, however,  inefficient on blockchains. This leads to the invention of the automated market maker, where traders, instead trading against other traders, only need to interact with a pool of assets reserved in a smart contract. Since the rise of DeFi, we have observed numerous such innovative DeFi design, most of which, however, have not been thoroughly studied. As a result, the risks and threats that DeFi users are exposed to are still unclear, necessitating empirical research to provide objective insights.

At the time of writing, Ethereum is the dominating permissionless blockchain hosting DeFi. The DeFi ecosystem on Ethereum reached a TVL of over $80$B~USD\footnote{In comparison, the Binance Smart Chain (BSC), ranked the second in terms of TVL at the time of writing, reaches~$20$B USD (cf.\ \url{https://debank.com/ranking/locked_value}). We omit BSC in this work because BSC starts to grow from early 2021, which has not accumulated sufficient data.}, with more than $50\%$ contributed by lending protocols. Lending and borrowing is a popular way to realize a leverage (amplifying the profit) in DeFi. A typical use case is outlined as follows. A trader collateralizes $\numprint{5000}$~USDT (a USD-pegged stablecoin, cf.\ Section~\ref{sec:stablecoinbackground}) to borrow~$1$ ETH, when the ETH/USDT price is $\numprint[ETH]{1} = \numprint[USDT]{3000}$. The borrower then sells the borrowed $\numprint[ETH]{1}$ for $\numprint[USDT]{3000}$. If the ETH price declines to, for example, $\numprint[ETH]{1} = \numprint[USDT]{2000}$, the trader can purchase~$1$ ETH with $\numprint[USDT]{2000}$, repay the debt, redeem the collateral, and finally realize a profit of $\numprint[USDT]{1000}$. The trader at the same time bears the liquidation risk if the ETH price increases and the USDT collateral is insufficient to back the $\numprint[ETH]{1}$ debt. In a liquidation, a liquidator repays the ETH debt for the trader and acquires the USDT collateral. The acquired collateral exceeds the rapid debt in value incurring a loss to the trader. Such repayment-acquisition liquidation mechanisms are adopted by most DeFi lending platforms. However, the incentives, risks (e.g., to what extend borrowers have lost in liquidation events), and stabilities of these protocols have not been thoroughly studied, which motivates this work.

We outline the details of the liquidation mechanisms in Section~\ref{sec:existing-protocols}. In the following, we introduce the essential components of DeFi that are relevant to lending protocols.

\subsubsection{Price Oracle}
Because lending protocols aim to liquidate collateralized assets upon collateral price declines, the lending smart contract is required to know the price of the collateral asset. Prices can either be provided through an on-chain oracle, such as smart contract based exchanges (e.g., Uniswap~\cite{uniswap2018}), or via an off-chain oracle (such as Chainlink~\cite{arijuel2017chainlink}). On-chain price oracles are known to be vulnerable to manipulation~\cite{qin2020attacking}.

\subsubsection{Flash Loan} The atomicity of blockchain transactions (executions in a transaction collectively succeed or fail) enables flash loans. A flash loan represents a loan that is taken and repaid within a single transaction~\cite{qin2020attacking,allen2020design}. A borrower is allowed to borrow up to all the available assets from a flash loan pool and execute arbitrary logic with the capital within a transaction. If the loan plus the required interests are not repaid, the whole transaction is reverted without incurring any state change on the underlying blockchain (i.e., the flash loan never happened). Flash loans are shown to be widely used in liquidations~\cite{qin2020attacking}.

\subsubsection{Stablecoin}\label{sec:stablecoinbackground}
Stablecoins are a class of cryptocurrencies designed to provide low price volatility~\cite{clark2020demystifying}. The price of a stablecoin is generally pegged to some reference point (e.g., USD). The typical stablecoin mechanisms are reserve of the pegged asset (e.g., USDT and USDC), loans (e.g., DAI), dual coin, and algorithmic supply adjustments~\cite{moin2020sok}.

\subsection{Terminology}\label{sec:terminology}
We adhere to the following terminologies in this paper.
\begin{description}
\item[Loan/Debt:] A borrower, secured by a collateral deposit, temporarily takes capital from a lender. The collateral is the insurance of the lender against defaults. 
\item[Interest Rate:] A loan is repaid by repaying the lent amount, plus a periodic percentage of the loan amount. The interest rate can be governed by the scarcity/surplus of the available asset supply within the lending smart contract.
\item[Over/Under-collateralization:] Blockchain based loans are typically over-collateralized, i.e., the borrower has to provide collateral assets of higher total value than the granted loan. A loan is under-collateralized when the value of the collateral is inferior to the debt.
\item[Position:] In this work, the collateral and debts are collectively referred to as a position. A position may consist of multiple-cryptocurrency collaterals and debts.
\item[Liquidation:] In the event of a negative price fluctuation of the debt collateral (i.e., a move below the liquidation threshold), a position can be liquidated. In permissionless blockchains, anyone can repay the debt and claim the collateral.
\item[Liquidation Threshold ($\mathbf{LT}$):] Is the percentage at which the collateral value is counted towards the borrowing capacity (cf. Equation~\ref{eq:borrowing-capacity}).
\item[Liquidation Spread ($\mathbf{LS}$):] Is the bonus, or discount, that a liquidator can collect when liquidating collateral (cf.\ Equation~\ref{eq:ls}). This spread incentivises liquidators to act promptly once a loan crosses the liquidation threshold.

\begin{equation}\label{eq:ls}
\begin{aligned}
    &Value\ of\ Collateral\ to\ Claim \\&= Value\ of\ Debt\ to\ Repay\times(1+\mathbf{LS})
\end{aligned}
\end{equation}
\item[Close Factor ($\mathbf{CF}$):] Is the maximum proportion of the debt that is allowed to be repaid in a single liquidation.
\item[Collateralization Ratio ($\mathsf{CR}$):] Is the ratio between the total value of collateral and debt (cf.\ Equation~\ref{eq:collateralizationratio}) where $i$ represents the index of collateral or debt if the borrower owns collateral or owes debt in multiple cryptocurrencies.

\begin{equation}\label{eq:collateralizationratio}
    \mathsf{CR}=\frac{\sum{Value\ of\ Collateral_i}}{\sum{Value\ of\ Debt_i}}
\end{equation}
A debt is under-collateralized if $\mathsf{CR}<1$, otherwise the debt is over-collateralized.
\item[Borrowing Capacity ($\mathsf{BC}$):] Refers to the total value that a borrower is allowed to request from a lending pool, given its collateral amount.
For each collateral asset $i$ of a borrower, its borrowing capacity is defined in Equation~\ref{eq:borrowing-capacity}.

\begin{equation}\label{eq:borrowing-capacity}
    \mathsf{BC} = \sum{Value\ of\ Collateral_i\times \mathbf{LT}_i}
\end{equation}
\item[Health Factor ($\mathsf{HF}$):] The health factor measures the collateralization status of a position, defined as the ratio of the borrowing capacity over the outstanding debts (cf.\ Equation~\ref{eq:health-factor}).

\begin{equation}\label{eq:health-factor}
\mathsf{HF}=\frac{\mathsf{BC}}{\sum{Value\ of\ Debt_i}}
\end{equation}
If $\mathsf{HF}<1$, the collateral becomes eligible for liquidation.
\end{description}

\section{Systematization of Lending and Liquidation Protocols}\label{sec:existing-protocols}
In this section, we systematize how the current borrowing mechanisms and their specific liquidation processes operate. 
\subsection{Borrowing and Lending System Model}\label{sec:borrowingandlendingsystemmodel}

\begin{figure}[tb!]
     \centering
    \includegraphics[width=0.9\columnwidth]{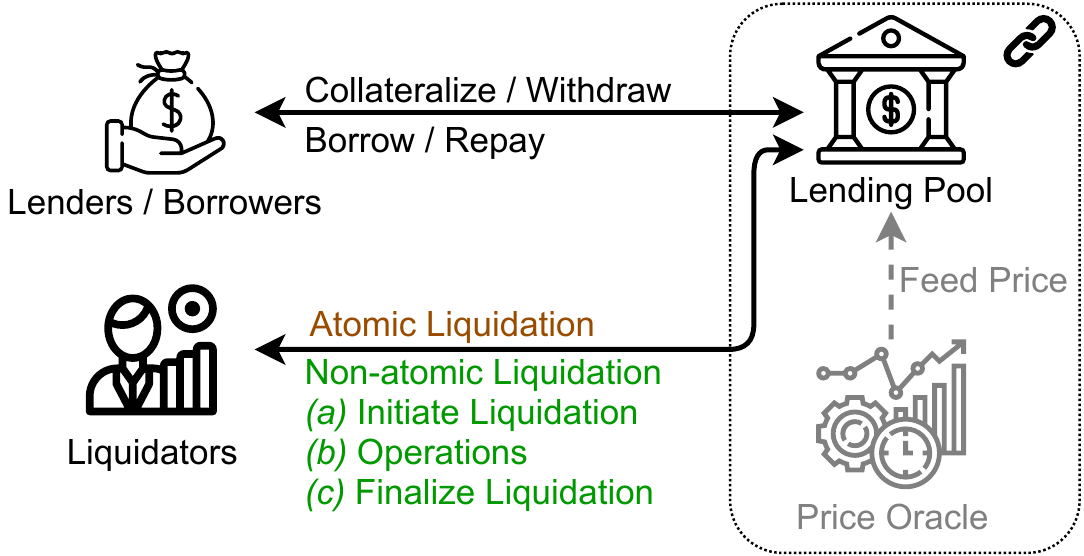}
    \caption{High-level system diagram of a lending pool system. 
    }
    \label{fig:liquidation_system_diagrams}
\end{figure}

The following aims to summarize the different actors engaging in borrowing and lending on a blockchain (cf.\ Figure~\ref{fig:liquidation_system_diagrams}).

A \textbf{lender} is an actor with surplus capital who would like to earn interest payments on its capital by lending funds to a third party, e.g., a borrower.

A \textbf{borrower} provides collateral to borrow assets from a lender. The borrower is liable to pay regular interest fees to the lender (typically measured as percentage of the loan). As lending in blockchains is typically performed without KYC, the borrower has to collateralize a value that is greater than the borrowed loan. In pure lending/borrowing platforms such as Aave, Compound and dYdX, the collateral from borrowers is also lent out as loans. Borrowers hence automatically act as lenders.

While lending could be performed directly on a peer-to-peer basis, blockchain-based lending protocols often introduce a \textbf{lending pool} governed by a smart contract. A pool can hold several cryptocurrencies, and users can interact with the pool to deposit or withdraw assets according to the rules defined by the smart contract code.

A \textbf{liquidator} observes the blockchain for unhealthy positions (i.e., the health factor is below~$1$) to be liquidated. Liquidators typically operate bots, i.e., automated tools which perform a blockchain lookup, price observation, and liquidation attempt, if deemed profitable. Liquidators are engaging in a competitive environment, where other liquidators may try to front-run each other~\cite{daian2019flash}. Notably, an atomic liquidation (e.g., a fixed spread liquidation) is settled in one blockchain transaction, while non-atomic liquidation mechanisms (e.g., auctions) generally require liquidators to interact with the lending pool in multiple transactions.


\subsection{Systematization of Liquidation Mechanisms}\label{sec:liquidationmechanisms}
We observe that existing lending platforms mainly adopt two distinct liquidation mechanisms. One mechanism is based on a non-atomic \emph{English auction}~\cite{krishna2009auction} process, and the other follows an atomic fixed spread strategy. We formalize the two existing dominating mechanisms as follows.


\subsubsection{Auction Liquidation}
An auction based liquidation mechanism follows the subsequent methodology:
\begin{enumerate}
    \item A loan becomes eligible for liquidation (i.e., the health factor drops below $1$).
    \item A liquidator starts the auction process (which can last several hours).
    \item Interested liquidators provide their bids (e.g., the highest bid receives the loan collateral).
    \item The auction ends according to the rules set forth in the auction contract.
\end{enumerate}

\begin{figure}[ht!]
    \centering
    \includegraphics[width=\columnwidth]{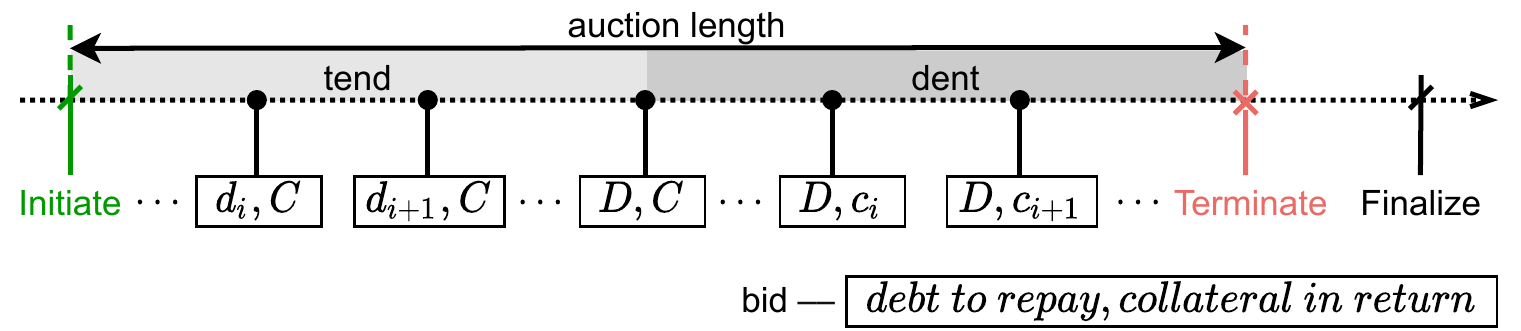}
    \caption{Two-phase liquidation auction of MakerDAO. Any actor can invoke the public function \emph{bite} to initiate the collateral auction and invoke the public function \emph{deal} to finalize the liquidation after the auction terminates.}
    \label{fig:maker-dao-auction}
\end{figure}


\paragraph{MakerDAO tend-dent auction} Specifically, MakerDAO employs a two-phase auction process, which we term \emph{tend-dent auction} (cf.\ Figure~\ref{fig:maker-dao-auction}). 
When a position with $D$ value of debt and $C$ value of collateral is eligible for liquidation, i.e., $\mathsf{HF}<1$, a liquidator is able to initiate the tend-dent auction. A liquidator is required to bid at a higher price than the last bid in the auction. The two-phase workflow is described as follows.
\begin{description}[leftmargin=10pt]
    \item[Tend:] In the \emph{tend} phase, liquidators compete by bidding to repay parts of the debt in exchange for the entire collateral. We denote the amount of debt committed to repay in each bid by $d_i$, s.t.~$d_i \leq D$ and $d_i>d_{i-1}$. If the auction terminates in the tend phase, the winning bidder receives all the collateral (i.e., $C$). When $d_i$ reaches $D$, the auction moves into the \emph{dent} phase.
    \item[Dent:] In the dent phase, liquidators compete by bidding to accept decreasing amounts of collateral in exchange for the full debt (i.e., $D$) they will end up repaying. We denote the amount of collateral committed in each bid by $c_i$, s.t.~$c_i \leq C$ and $c_i<c_{i-1}$. The winning bidder repays the full debt and receives the partial collateral (denoted by $c_{\text{win}}$). The remaining collateral (i.e., $C-c_{\text{win}}$) is returned to the position owner (i.e., the borrower).
\end{description}

The auction terminates when any of the following two conditions is satisfied. Note that the auction can terminate in the tend phase.
\begin{enumerate}
    \item \textbf{Auction Length Condition:} the configurable auction length (e.g., $6$ hours) has passed since the initiation of the auction.
    \item \textbf{Bid Duration Condition:} the configurable bid duration (e.g., $5$ hours) has passed since the last bid.
\end{enumerate}

After the termination of an auction, the winning liquidator is allowed to finalize the liquidation to claim the proposed collateral.

\subsubsection{Fixed Spread Liquidation}
Instead of allowing multiple liquidators to bid over a time-frame, a liquidatable loan can be instantly liquidated with a pre-determined discount (cf.\ Figure~\ref{fig:liquidation_system_diagrams}). Aave, for instance, allows liquidators to purchase the loan collateral at up to a 15\% discount of the current market price. This discount, or liquidation spread, is known upfront, and the liquidators can hence locally decide whether to engage in a liquidation opportunity.
Following a fixed spread model avoids hour-long liquidation auctions, which cost time and transaction fees. Liquidators, moreover, can choose to liquidate collateral with the use of atomic flash loans~\cite{qin2020attacking}. While flash loans increase the transaction costs of the liquidators, they reduce the currency exposure risk of holding the assets required for liquidation.

\paragraph{Fixed Spread Liquidation Example}
In the following, we provide an example of a fixed spread liquidation:
\begin{enumerate}
    \item \textbf{Currency values:} We assume an initial price of $\numprint{3500}$ USD/ETH.
    \item \textbf{Collateral Deposit:} A user deposits $3$ ETH, and hence has $\numprint[USD]{10500}$ worth of collateral. If we assume a liquidation threshold ($\mathbf{LT}$, cf.\ Section~\ref{sec:terminology}) of $0.8$, the resulting borrowing capacity of the user is $\mathsf{BC} = \numprint[USD]{10500}\times\mathbf{LT} = \numprint[USD]{8400}$.
    \item \textbf{Borrowing:} In the next step the user borrows, for instance, $\numprint[USDC]{8400}$ worth $\numprint[USD]{8400}$.
    \item \textbf{ETH price decline:} We now assume that the ETH value declines to $\numprint[USD/ETH]{3300}$, which means that the collateral value declines to $\numprint[USD]{9900}$ with $\mathsf{BC}=\numprint[USD]{7920}$. The price oracle updates the ETH price on the lending smart contract. The health factor of the loan now drops to $\mathsf{HF}=\frac{\numprint[USD]{7920}}{\numprint[USD]{8400}}\approx 0.94 < 1$ and thus the collateral is available for liquidation.
    \item \textbf{Liquidation:} A liquidator submits a liquidation transaction to repay $50$\% (close factor $\mathbf{CF}$, cf.\ Section~\ref{sec:terminology}) of the debt, i.e., $\numprint[USDC]{4200}$. In return, the liquidator is allowed to purchase collateral at the price of $\frac{\numprint[USD/ETH]{3300}}{1+\mathbf{LS}}=\numprint[USD/ETH]{3000}$ (we assume that the liquidation spread $\mathbf{LS}$ is $10\%$, cf.\ Section~\ref{sec:terminology}). 
    In this liquidation, the liquidator receives $\frac{\numprint[USD]{4200}}{\numprint[USD/ETH]{3000}}\times\numprint{3300}$ USD/ETH $=\numprint[USD]{4620}$ worth of ETH and realizes a profit of $\numprint[USD]{420}$.
    
\end{enumerate}

\subsection{Studied Lending Protocols}
Within this work we focus on the biggest lending protocols, measured by total value locked, notably MakerDAO ($12.49$B~USD), Aave ($11.20$B~USD), Compound ($10.15$B~USD), and dYdX ($247.6$M~USD).

Aave is a pool-based lending and borrowing protocol~\cite{aave}. Lenders deposit assets into a pool governed by open-source smart contracts, and borrowers can then take loans out of this pool. The interest rate of an Aave pool is decided algorithmically by the smart contract and depends on the available funds within the lending pool. The more users borrow an asset, the higher its interest rate rises. A lending pool can consist of several cryptocurrency assets, for instance ETH, DAI, and USDC.
In Aave, when the health factor drops below~$1$, any liquidator can call the public pool function \emph{liquidationCall}, by repaying parts or all of the outstanding debt, while profiting from the liquidation spread. Aave specifies that only a maximum of $50\%$ of the debt can be liquidated within one \emph{liquidationCall} execution (referred to as a close factor). The liquidation spread on Aave ranges from 5\% to 15\%, depending on the considered markets. Aave bases its pricing feed on the external Chainlink oracle~\cite{arijuel2017chainlink}. Aave was upgraded to a newer version in December 2020 while the core protocols remained nearly unchanged. In this work, we distinguish the two versions with Aave V1 and Aave V2.

Compound~\cite{compoundfinance} launched before Aave and operates in a similar fashion. Users deposit assets and earn interests based on the amount of interests paid by borrowers. When a borrower exceeds the borrowing capacity, at most 50\% of outstanding debt can be repaid at once by a liquidator (same as the close factor in Aave). The liquidation may continue until the collateral guarantees a health factor superior to one. The liquidator in exchange receives the collateral at the current market price minus the liquidation spread.

dYdX~\cite{dydx} is divided into two sub-protocols, one for trading, borrowing, lending and one that also supports futures markets. Similar to Aave and Compound, dYdX operates at a fixed spread of $5\%$ for the \wrapletters{WETH/USDC}, \wrapletters{WETH/DAI} and \wrapletters{USDC/DAI} markets, at the time of writing. dYdX's close factor is $100$\%, allowing the liquidators to liquidate the entire collateral within one liquidation.

Contrary to the aforementioned borrowing/lending protocols, MakerDAO provides a decentralized stablecoin called DAI that is pegged to the US dollar, while still functioning financially similar to a borrowing/lending platform. A user can collateralize at least, e.g., $150$\% of a crypto asset (for instance ETH) to mint $100$\% DAI. Creating DAI opens a so-called \emph{collateralized debt position (CDP)}, which can be liquidated if the collateralized value drops below the fixed collateralization ratio. MakerDAO adopts an auction-based liquidation mechanism (cf.\ Section~\ref{sec:liquidationmechanisms}).

\section{Liquidation Insights}\label{sec:insights}
By observing the publicly readable Ethereum blockchain we compiled the following insights on liquidation events.
\subsection{Measurement Setup}
\begin{figure}[b]
    \centering
    \includegraphics[width=\columnwidth]{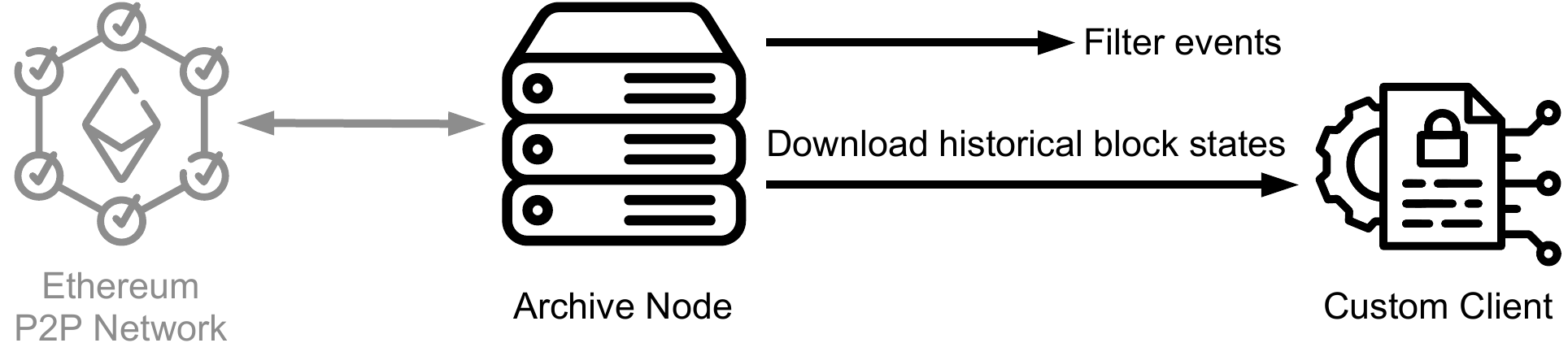}
    \caption{Measurement setup overview.}
    \label{fig:setup-overview}
\end{figure}
Our measurement setup is outlined in Figure~\ref{fig:setup-overview}. We gather our data by crawling blockchain events (e.g., liquidation events) and reading blockchain states (e.g., oracle prices) from an Ethereum full archive node, on an AMD Ryzen Threadripper 3990X with $64$ cores, $256$ GB of RAM and $2\times8$ TB NVMe SSD in Raid $0$ configuration. An Ethereum archive node stores not only the blockchain data but also the chain state at every historical block, which supports efficient historical state query (e.g., the borrowing position debt amount at a specific block). At the time of writing, an Ethereum archive node requires over $4$ TB disk space.

The Ethereum events are essentially EVM logs, which are also recorded on-chain. An event is indexed by its signature, a 256-bit hash, and the contract address emitting this event. We hence can filter the liquidation events emitted from the studied lending pools.

We also build our own custom Ethereum client based on the golang-based geth client\footnote{\url{https://github.com/ethereum/go-ethereum}} to execute transactions on a specific block (the block state is downloaded from the archive node) when necessary. For instance, in Section~\ref{sec:optimalstrategy}, we validate our optimal fixed spread liquidation strategy through concrete executions on past blockchain states.

\subsection{Overall Statistics}
\begin{figure}[tb!]
    \centering
    \includegraphics[width=\columnwidth]{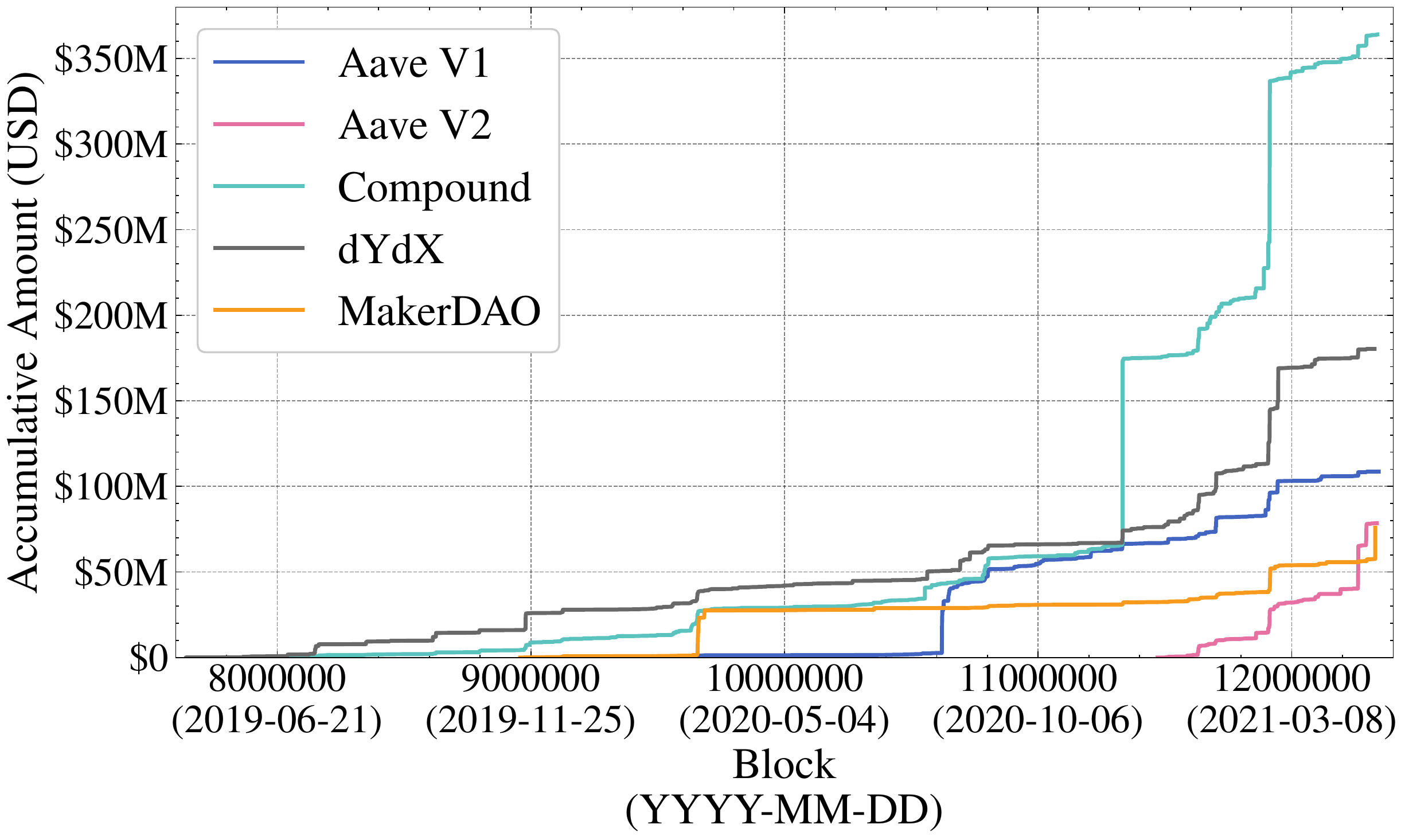}
    \caption{From the inception of each considered lending protocol, we plot the accumulative collateral sold through liquidation (April,~2019~to April,~2021).}
    \label{fig:accumulative_liquidation_stacked}
\end{figure}

\begin{figure*}[tb]
    \centering
    \includegraphics[width=\textwidth]{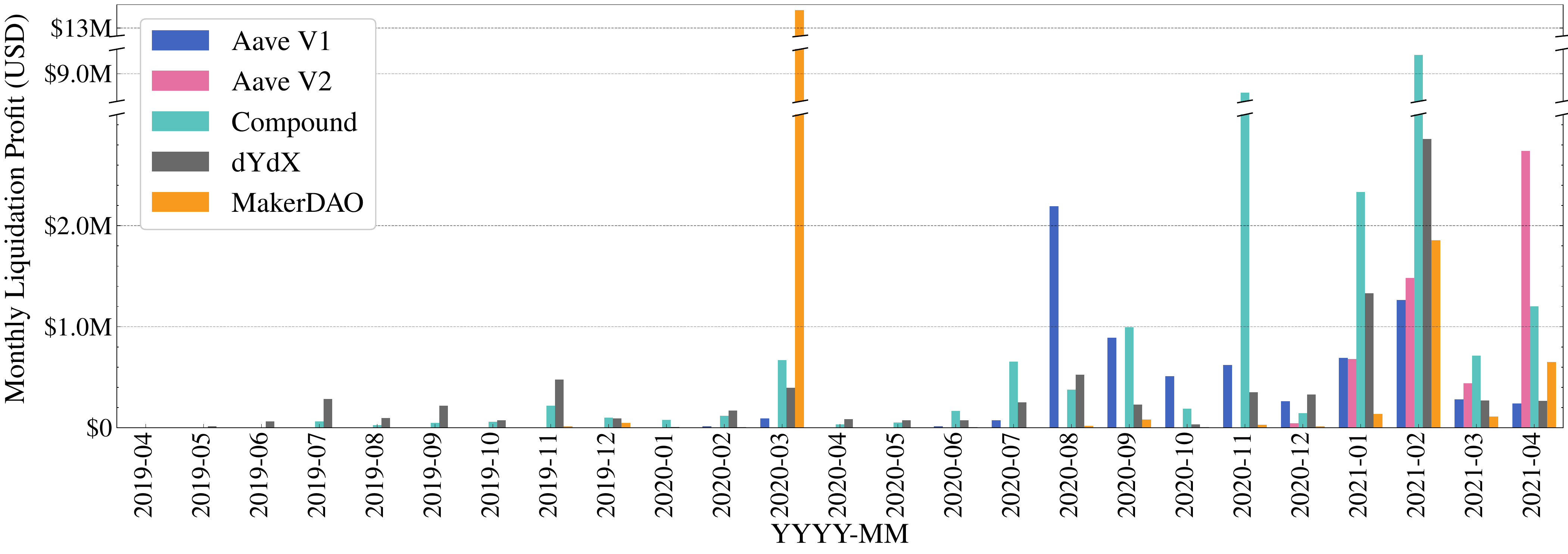}
    \caption{Monthly accumulated liquidator profit. We observe an outlier for MakerDAO in March,~2020, because the MakerDAO liquidation bots were faulty due to an excessive price decline of ETH. The outlier for Compound in November,~2020~is caused by an irregular price reported by a price oracle.}
    \label{fig:monthly-liquidation-profit}
\end{figure*}

In total, we observe \TotalSuccessfulLiquidations successful liquidations from the inception\footnote{The inception blocks of Aave, Compound, dYdX and MakerDAO are~\block{9241022}, \block{7710733}, \block{7575711} and \block{8040587} respectively.} of the four platforms to block \CrawlingEndBlock, the last block in the month of April,~2021. We normalize the values of different cryptocurrencies to USD according to the prices given by the platforms' on-chain price oracles at the block when the liquidation is settled. We crawl this data on-chain, and do not rely on an external price oracle, or API.

In Figure~\ref{fig:accumulative_liquidation_stacked}, we present the accumulative collateral sold through liquidation in terms of USD. The overall liquidated collateral on the four platforms Aave, Compound, dYdX, and MakerDAO accumulates to a total of~\AccumulativeLiquidatedAmount. We notice an increase on Compound in November,~2020. This is caused by an irregular DAI price provided by the Compound price oracle, which triggers a large volume of cryptocurrencies to be liquidated~\cite{OracleEx18:online}. Another remarkable boost on Compound in February,~2021 is caused by the drastic fluctuations in prices of cryptocurrencies~\cite{Over117M34:online}.

\subsection{Incentives and Participation}
In the following, we measure how liquidators are incentivized to engage in liquidations and elaborate on the status quo of liquidator participation.

\subsubsection{Liquidator Profit \& Loss}\label{sec:profitandloss}
To measure the profit of each liquidation event, we assume that the purchased collateral is immediately sold by the liquidator at the price given by the price oracle. The total profit by the \TotalSuccessfulLiquidations liquidations sums up to a total of \TotalLiquidationProfit. To better understand the temporal evolution of liquidation profits, we show the monthly collective profit yielded from each platform in Figure~\ref{fig:monthly-liquidation-profit}.

MakerDAO notably shows an outlier in March,~2020, when the MarkerDAO monthly profit reached~\MakerDAOMarchProfit. This outlier is due to the Ethereum network congestion caused by the~$43$\% ETH price market collapse on the~13th March,~2020~\cite{maker-fail}. The liquidation bots were not acting accordingly, which caused the liquidation transactions to not be swiftly included in the blockchain. This delay allowed other capable liquidators to manually win the auctions at a negligible cost.

In November,~2020, an irregular DAI price provided by the Compound price oracle~\cite{OracleEx18:online} allowed liquidators to profit in total~\ComopoundNovemberProfit. We also observe that Compound contributes a strong liquidation profit of \ComopoundFebruaryProfit in February,~2021, which, however, does not seem related to any bot failure or oracle irregularity.

To study the number of liquidators, we assume that each unique Ethereum address represents one liquidator. We then identify a total of \TotalNumLiquidators unique liquidators. On average the liquidators yield a profit of \TotalPerLiquidatorProfit each. We show the number of liquidators and their average profit on the four considered platforms in Table~\ref{tab:liquidator_profit}. Remarkably, the most active liquidator performs \MostActiveLiquidatorLiquidations liquidations alone, which yield a total profit of \MostActiveLiquidatorProfit. The most profitable liquidator generates \MostLucrativeLiquidatorProfit in only \MostLucrativeLiquidatorLiquidations liquidations.

\begin{table}[bt!]
    \centering
    \caption{Number of the liquidations and liquidators on Aave, Compound, dYdX, MakerDAO and their average profit. We measure the number of liquidators based on their unique Ethereum address. We notice that some liquidators operate on multiple lending markets.
    }
    \resizebox{\columnwidth}{!}{%
    \begin{tabular}{lrrr}
    \toprule
    \bf Platform & \bf Liquidations  & \bf Liquidators & \bf Average Profit \\
    \midrule
    Aave V1 & \AaveLiquidations & \AaveNumLiquidators  & \AavePerLiquidatorProfit \\
    Aave V2 & \AaveVTLiquidations & \AaveVTNumLiquidators  & \AaveVTPerLiquidatorProfit \\
    Compound & \CompoundLiquidations & \CompoundNumLiquidators  & \CompoundPerLiquidatorProfit \\
    dYdX & \dYdXLiquidations & \dYdXNumLiquidators & \dYdXPerLiquidatorProfit \\
    MakerDAO & \MakerDAOAuctions & \MakerDAONumLiquidators  & \MakerDAOPerLiquidatorProfit \\
    \midrule
    Total& \TotalSuccessfulLiquidations & \TotalNumLiquidators &  \TotalPerLiquidatorProfit \\
    \bottomrule
    \end{tabular}%
    }
    \label{tab:liquidator_profit}
\end{table}

We also discover~\NonProfitableMakerDAOLiquidations MakerDAO liquidations that are not profitable and incur a total loss of \NonProfitableMakerDAOLoss. After manually inspecting those non-profitable liquidation transactions, we can confirm that the liquidation losses are caused by collateral price fluctuations during the auctions.

\begin{figure*}[tb!]
    \centering
    \includegraphics[width=\textwidth]{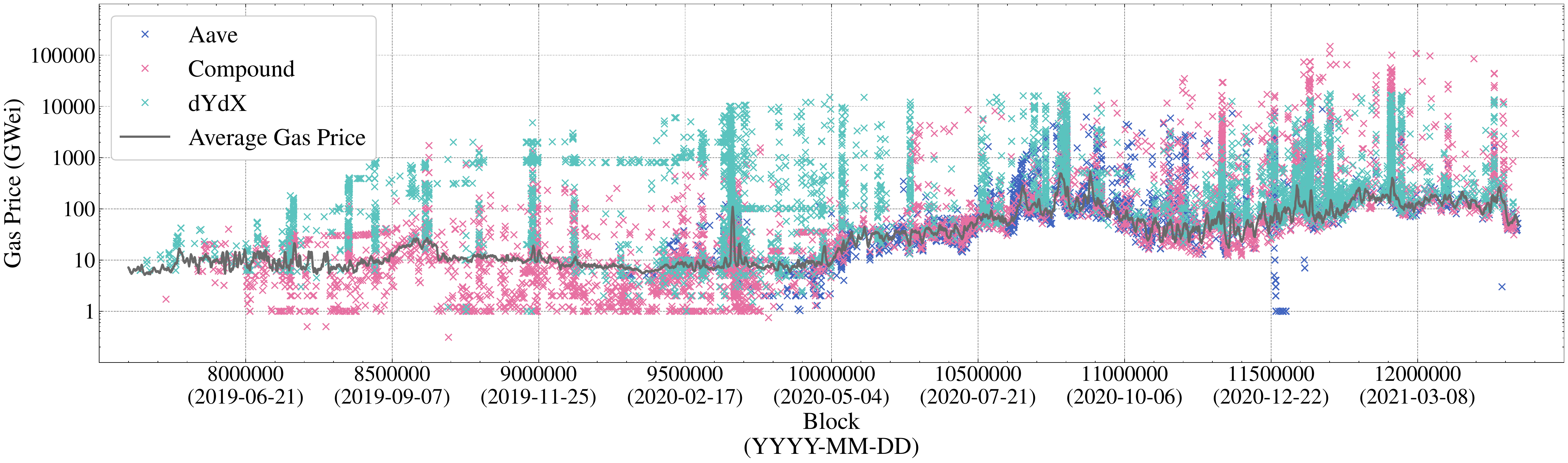}
    \caption{Gas prices paid by liquidators. We also report the 6000-blocks (1 day) moving average of the block median measured on-chain. Interestingly, several liquidations are below the average gas price. We notice a gas price spike in March,~2020~because of a collapse of ETH price~\cite{maker-fail}. There is an uptrend of gas price since May,~2020~due to the growing popularity of DeFi.}
    \label{fig:liquidation-gas_prices}
\end{figure*}

\subsubsection{Fixed Spread Liquidations}\label{sec:fixedspreadparticipation}
We observe \AaveLiquidations, \AaveVTLiquidations, \CompoundLiquidations and \dYdXLiquidations settled liquidations on Aave~V1, Aave~V2, Compound and dYdX respectively.
\paragraph{Liquidator Participation} In Figure~\ref{fig:liquidation-gas_prices}, we show the gas price of every fixed spread liquidation transaction along with the average gas price. Note that we show the 6000-block moving average of the block gas price medians in the figure to smooth the curve for readability. The data in Figure~\ref{fig:liquidation-gas_prices} shows that many liquidators pay significant gas fees (the y-axis is a log scale). We find that~\NumLiquidationsHigherThanAverageGasPrice of the liquidations pay an above average transaction fee, and hence allows the conclusion that liquidation events are competitive.

\subsubsection{Auction Liquidations}\label{sec:auctionparticipation}

Out of the recorded~\MakerDAOAuctions MakerDAO liquidations,~\MakerDAOTerminateInTend auctions terminate in the tend phase and the other~\MakerDAOTerminateInDent auctions terminate in the dent phase. The average number of bidders participating in a liquidation is only~\MakerDAOAverageBidders.
We notice that~\MakerDAOBidsPerAuction bids (\MakerDAOTendBidsPerAuction tend bids and~\MakerDAODentBidsPerAuction dent bids), are placed per auction.

\begin{figure}[tb!]
    \centering
    \includegraphics[width=\columnwidth]{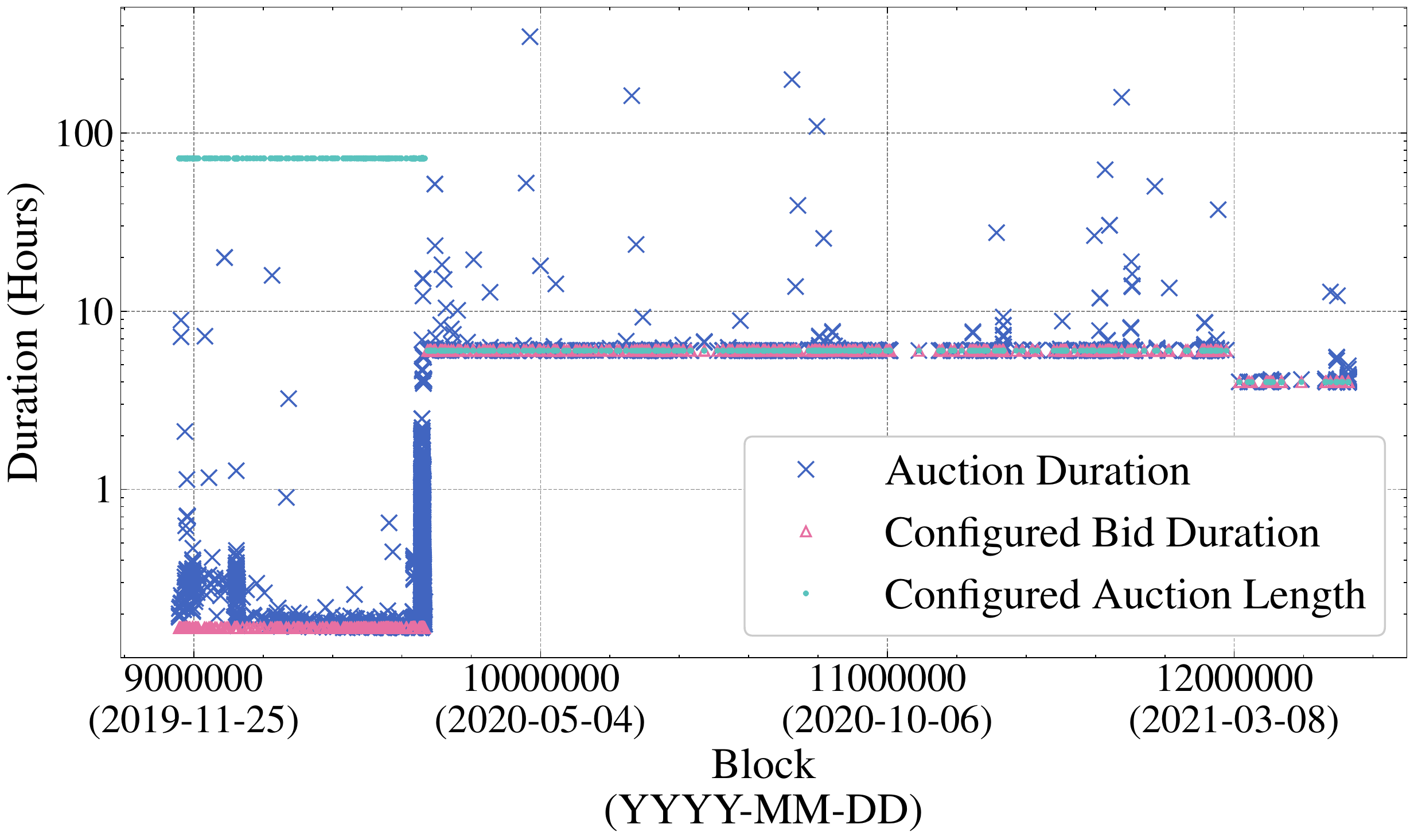}
    \caption{Duration of the MakerDAO liquidations. Interestingly, auctions last longer than they're configured to. We can observe a change in system auction parameters after~the 13th March,~2020~event (the MakerDAO liquidation incident).}
    \label{fig:makerdao-auction-duration}
\end{figure}

\paragraph{Duration} We define the duration of a MakerDAO liquidation auction, as the time difference between the auction initiation and finalization. To capture time, we resort to the block timestamps. We visualize the duration of the MakerDAO liquidations in Figure~\ref{fig:makerdao-auction-duration}. On average, an liquidation lasts for~\MakerDAOAverageAuctionDuration (mean$\pm$standard deviation).
There are~\MakerDAOAuctionsInOneHour auctions terminating within one hour. We observe that few liquidations last longer than intended, which can be explained by that fact that the respective liquidators did not finalize the auction and hence didn't claim the liquidation proceeds. For example, the longest auction lasts for~\LongestMakerDAOAuction, while its last bid is placed~\LongestMakerDAOAuctionLastBid prior to the termination.

\paragraph{Bid Intervals} To better understand the bidding process of the various liquidators, we study the number of bids, and their respective intervals. We note that the first bid is placed on average~\MakerDAOFirstBidInterval after the auction initiation. Given that an auction can last several hours, it appears that most bidders engage early in the auction process. We observe that \MakerDAOAuctionsWithMoreThanOneBids auctions terminate with more than one bid placed. It also appears that bids come close together, as the average interval between bids is \MakerDAOBidInterval.

\subsection{Risks}
We proceed to discuss the risks that the participants of a lending pool (i.e., borrowers, lenders and liquidators) bear in liquidations.
\subsubsection{The Problem of Over-Liquidation}\label{sec:overliquidation}
We observe that the liquidation mechanisms of Aave and Compound grant a liquidator the right to liquidate up to~$50$\% of the collateral (i.e., the close factor) once a debt becomes liquidatable. dYdX even allows a liquidator to purchase~$100$\% of the collateral at the fixed spread discount.


Such design decision favor the liquidators over the borrowers, as a debt can likely be rescued by selling less than $50$\% of its value. Choosing an appropriate close factor is challenging, because the liquidation mechanism should minimize the number of liquidation events and overall transactions due to the limited transaction throughput of blockchains.

Auction mechanisms do not specify a close factor and hence offer a more granular method to liquidate collateral. In return, auction liquidators are exposed to the risk of loss due to the price fluctuations of the collateral during the liquidation (cf.\ Appendix~\ref{app:postliquidationprice}). The MakerDAO tend-dent auction, is to the best of our knowledge, the only auction mechanism that is widely adopted in blockchain liquidations. Yet, the liquidation bots must remain robust and vigilant under blockchain congestion, otherwise the borrowers may endure excessive losses~\cite{maker-fail}. Some alternative auction mechanisms (e.g., Vickrey auction~\cite{ausubel2006lovely} and Dutch reverse auctions~\cite{horlen2005reverse}) may have potential in mitigating the over-liquidation problem and could prove more resilient to network congestion. We leave an objective comparison of different auction designs for future work.

\subsubsection{Bad Debts}
\label{sec:bad_debts}

We define a borrowing position as a bad debt if it is financially rationale for neither the borrowers nor the lending platform to close the position. In the following we introduce two types of bad debts.

\begin{enumerate}
\item \textbf{Type I bad debt (Under-collateralized position):} If the collateral value falls below the value of the debt, then either the borrower or the lending platform will suffer a loss if the corresponding position is closed. Type I bad debt is typically caused by overdue liquidations. An overdue liquidation can, e.g., occur when the collateral/debt asset suffers a severe price fluctuation or the blockchain network is congested.
\item \textbf{Type II bad debt (Excessive Transaction Fees):} When an over-collateralized position is closed, the borrower will regain the excess asset used for over-collateralization. However, if the value of the excess asset cannot cover the transaction fee, then there is no incentive for the borrower to repay and close this position.
\end{enumerate}

Because the value of the debt plus transaction fee is superior to the value of the bad debt collateral, repaying bad debt is not a financially rational endeavor for a borrower. The accumulation of bad debt reduces the total liquidity in a lending protocol, which necessarily leads to higher interest rates for borrowers. If the lending pool maintains exclusively bad debts, lenders will not be able to withdraw funds.

In the following we quantitatively measure the amount of bad debts present in existing lending protocols. To that end, we first have to assume a somewhat random cost which a borrower would need to bear, when repaying debt. For the sake of the example here, we choose a cost of~$100$ USD to repay the debt, and consider the blockchain state at block~\block{12344944} (30th~Apr~2021). Given this cost, we have identified in total~$351$/$\numprint{3525}$ Type I/II bad debts (cf.\ Table~\ref{tab:bad_debt}). Remarkably, the liquidity of Aave V2 is reduced by~$87.4$K~USD due to existing bad debts. It is worth mentioning that dYdX does not have any Type I bad debt at block~\block{12344944}. This is, because dYdX apparently uses an external insurance fund, to write off bad debts of Type I.

\begin{table}[bt!]
    \centering
    \caption{Statistics of Type I/II bad debts on Aave, Compound and dYdX at block~\block{12344944} (30th~Apr~2021). For instance, if it costs~$100$ USD for a borrower to repay its debt, then~$2,550$ ($32.0\%$) of the lending position are classified as Type II bad debts on Compound, which causes 14.3K USD collateral value to be locked.}
    \resizebox{\columnwidth}{!}{%
    \begin{tabular}{l|c|cc}
    \toprule
            & Type I                                                                     & \multicolumn{2}{c}{Type II}                                                                                                                               \\
    Transaction fee & -                                                                          & $\leq$ 10 USD                                                              & $\leq$ 100 USD                                                               \\
    \midrule
    Aave V2         & \begin{tabular}[c]{@{}l@{}}28 (0.5\%)\\ \numprint{25379} USD collateral\end{tabular}  & \begin{tabular}[c]{@{}l@{}}102 (1.9\%)\\ \numprint{4793} USD collateral\end{tabular}  & \begin{tabular}[c]{@{}l@{}}255 (4.7\%)\\ \numprint{62017} USD collateral\end{tabular}   \\\midrule
    Compound        & \begin{tabular}[c]{@{}l@{}}333 (4.2\%)\\ \numprint{27473} USD collateral\end{tabular} & \begin{tabular}[c]{@{}l@{}}\numprint{1681} (21.1\%)\\ 675 USD collateral\end{tabular} & \begin{tabular}[c]{@{}l@{}}\numprint{2550} (32.0\%)\\ \numprint{14399} USD collateral\end{tabular} \\\midrule
    dYdX            & -                                                                          & \begin{tabular}[c]{@{}l@{}}411 (36.3\%)\\ \numprint{1287} USD collateral\end{tabular} & \begin{tabular}[c]{@{}l@{}}720 (63.5\%)\\ \numprint{18019} USD collateral\end{tabular}  \\
    \bottomrule
    \end{tabular}
    }
    \label{tab:bad_debt}
\end{table}

\begin{table}[bt]
 \centering
    \caption{Statistics of unprofitable liquidation opportunities on Aave, Compound and dYdX at block~\block{12344944} (30th~Apr~2021). For instance, Aave configures a $50\%$ liquidation threshold with at most $15\%$ liquidation spread. Based on our measurement, at least $59.1\%$ of the Aave liquidation opportunities are not profitable if the liquidation process costs~$100$.}
    \resizebox{\columnwidth}{!}{%
        \begin{tabular}{llll}
        \toprule
        Transaction Fee &                            & $\leq$ 10 USD       & $\leq$ 100 USD      \\
        \midrule
        Aave V2         & $\mathbf{LT}=50\%, \mathbf{LS} \leq 15\%$        & \begin{tabular}[c]{@{}l@{}}$\geq$ 6 (27.2\%)\\ \numprint{398} USD collateral\end{tabular}  & \begin{tabular}[c]{@{}l@{}}$\geq$ 13 (59.1\%)\\ \numprint{3404} USD collateral\end{tabular}  \\
        Compound        & $\mathbf{LT}=50\%, \mathbf{LS} = 8\%$            & \begin{tabular}[c]{@{}l@{}}325 (14.8\%)\\ \numprint{34025} USD collateral\end{tabular}         & \begin{tabular}[c]{@{}l@{}}350 (15.9\%)\\ \numprint{125722} USD collateral\end{tabular}  \\
        dYdX            & $\mathbf{LT}=100\%, \mathbf{LS} = 5\%$           & -                   & -            \\
        \bottomrule
        \end{tabular}
    }
    \label{tab:unprofitable_liquidation}
\end{table}

\subsubsection{Unprofitable Liquidations}\label{sec:unprofitable_liquidation}

We define a liquidation opportunity as unprofitable if the bonus collected by the liquidator cannot cover the transaction fee. Unprofitable liquidations imply that the lending position's health cannot be restored in time, which necessarily leads to an accumulation of Type I bad debt. We measure the number of unprofitable liquidation opportunities in Table~\ref{tab:unprofitable_liquidation}. Given the average transaction fee at the time of writing, we choose a transaction cost of~$100$ USD for a single liquidation. Remarkably, by block~\block{12344944} (30th~Apr~2021), we have identified~$350$ unprofitable liquidation opportunities on Compound, which corresponds to \numprint{125722} USD worth of collateral. Rational liquidators will not attempt to liquidate unprofitable lending positions. Therefore, these positions will inevitably become Type I bad debts, if their health factor continues to fall.

\subsubsection{Flash Loan Usages}\label{sec:flash_loan_usages}
Fixed spread liquidations can be conducted in one transaction alone, which reduces the complexity and risk for the liquidators. For instance, liquidators do not need to hold any assets ready to repay debt for a liquidation. Instead, the liquidators can resort to flash loan pools, lend the required capital to repay debt, and pay back the flash loan interests by the end of the liquidation call. A typical flow of a flash loan liquidation works as follows:
\begin{enumerate}
    \item The liquidator borrows a flash loan in currency $X$ to repay the debt.
    \item The liquidator repays the borrower's debt with a flash loan and receives collateral in currency $Y$ at a premium.
    \item The liquidator exchanges parts of the purchased collateral in exchange for currency $X$.
    \item To conclude the flash loan, the liquidator repays the flash loan together with flash loan interest. The remaining profit lies with the liquidator. If the liquidation is not profitable, the flash loan would not succeed.
\end{enumerate}

To understand to what degree liquidators engage in flash loans, we study the flash loan pools of Aave and dYdX (which also act as lending pools). We therefore filter the relevant events in the liquidation transactions that apply to flash loans. For our observation window, we observe a total of~\TotalFlashLoans flash loans, that are borrowed for liquidations. The accumulative flash loan amount lent sums up to~\TotalFlashLoanAmount. We summarize further details in Table~\ref{tab:flash_loan_usages}. In the table, we include the flash loans that are borrowed before and repaid after liquidation. We notice that in terms of accumulative amounts, dYdX flash loans are more popular than Aave, likely due to the low interest rate of dYdX flash loans.

\begin{table}[tb!]
\centering
\caption{Flash loan usages for liquidations.}
\resizebox{\columnwidth}{!}{%
\begin{tabular}{@{}cccr@{}}
\toprule
\multicolumn{1}{c}{\textbf{\begin{tabular}[c]{@{}c@{}}Liquidation\\ Platform\end{tabular}}} & \multicolumn{1}{c}{\textbf{\begin{tabular}[c]{@{}c@{}}Flash Loan\\ Platform\end{tabular}}} & \textbf{Flash Loans} & \multicolumn{1}{c}{\textbf{\begin{tabular}[c]{@{}c@{}}Accumulative\\ Amount\end{tabular}}} \\ \midrule
Aave V1 & dYdX & \AavedYdXFlashLoans & \AavedYdXFlashLoanAmount \\ \midrule

\multirow{2}{*}{Aave V2} & Aave V2 & \AaveVTAaveVTFlashLoans & \AaveVTAaveVTlashLoanAmount \\ \cmidrule(l){2-4} 
 & dYdX & \AaveVTdYdXFlashLoans & \AaveVTdYdXlashLoanAmount \\ \midrule

\multirow{2}{*}{Compound} & Aave V1 & \CompoundAaveFlashLoans & \CompoundAaveFlashLoanAmount \\ \cmidrule(l){2-4} 
 & dYdX & \CompounddYdXFlashLoans & \CompounddYdXFlashLoanAmount \\ 
\bottomrule
\end{tabular}%
}
\label{tab:flash_loan_usages}
\end{table}

\subsection{Instabilities}
In this section, we discuss the lending platform instabilities due to cryptocurrency price fluctuations.
\subsubsection{Liquidation Sensitivity}\label{sec:sensitivity}
To understand how the lending platforms respond to price declines of different currencies, we quantify the liquidation sensitivity, i.e., the amount of collateral that would be liquidated, if the price of the collateral would decline by up to $100$\%. We again capture Aave V2, Compound, MakerDAO, and dYdX in the snapshot state at block~\CrawlingEndBlock\footnote{We ignore Aave V1 in the sensitivity measurement because the majority of the liquidity of Aave V1 had been migrated to Aave V2 at block~\CrawlingEndBlock.} to provide an exhaustive understanding of borrower risk profiles.

\newcommand\mycommfont[1]{\ttfamily\textcolor{blue}{#1}}
\SetCommentSty{mycommfont}
\SetAlFnt{\small}
\SetAlCapFnt{\small}
\SetAlCapNameFnt{\small}
\begin{algorithm}[t]
\SetAlgoLined
\SetKwProg{Fn}{Function}{:}{end}
\SetKwInOut{Input}{Input}\SetKwInOut{Output}{Output}
\SetKwFunction{Collateral}{Col}\SetKwFunction{Debt}{Debt}

\Input{Target currency $\mathfrak{C}$; Price decline percentage $d\%$; The set of borrowers $\{\mathcal{B}_i\}$\;}
\Output{Liquidatable collateral $\mathsf{LC}$\;}
\BlankLine
\Fn{\Collateral{$\mathcal{B}$, $c$}}{
    \KwRet{The value of collateral in currency $c$ owned by $\mathcal{B}$\;}
}
\BlankLine
\Fn{\Debt{$\mathcal{B}$, $c$}}{
\KwRet{The value of debt in currency $c$ owed by $\mathcal{B}$\;}}
\BlankLine
$\mathsf{LC} \leftarrow 0$\;
\ForEach{$\mathcal{B}\in\{\mathcal{B}_i\}$}{
\If{$\mathcal{B}$ owns collateral in currency $\mathfrak{C}$}{

\tcp*[h]{The collateral value of $\mathcal{B}$}

\tcp*[h]{after the price decline}

$\mathsf{C}\leftarrow\sum_c$\Collateral{$\mathcal{B}$, $c$}$-$\Collateral{$\mathcal{B}$, $\mathfrak{C}$}$\times d\%$\;

\tcp*[h]{The borrowing capacity of $\mathcal{B}$}

\tcp*[h]{after the price decline}

$\mathsf{BC}\leftarrow\sum_c$\Collateral{$\mathcal{B}$, $c$}$\times \mathbf{LT}_c - $\Collateral{$\mathcal{B}$, $\mathfrak{C}$}$\times \mathbf{LT}_\mathfrak{C}\times d\%$\;

\tcp*[h]{The debt value of $\mathcal{B}$} 

\tcp*[h]{after the price decline}

$\mathsf{D}\leftarrow\sum_c$\Debt{$\mathcal{B}$, $c$}\;

\If{$\mathcal{B}$ owes debt in currency $\mathfrak{C}$}{
$\mathsf{D}\leftarrow\mathsf{D}-$\Debt{$\mathcal{B}$, $\mathfrak{C}$}$\times d\%$\;
}
\If{$\mathsf{BC}<\mathsf{D}$}{
$\mathsf{LC} \leftarrow \mathsf{LC} + \mathsf{C}$\;
}
}
}
\KwRet{$\mathsf{LC}$\;}
\caption{Sensitivity measurement algorithm.}
\label{alg:liquidationsensitivity}
\end{algorithm}

\begin{figure*}[tb]
     \centering
     \begin{subfigure}[b]{0.495\textwidth}
         \centering
         \includegraphics[width=\textwidth]{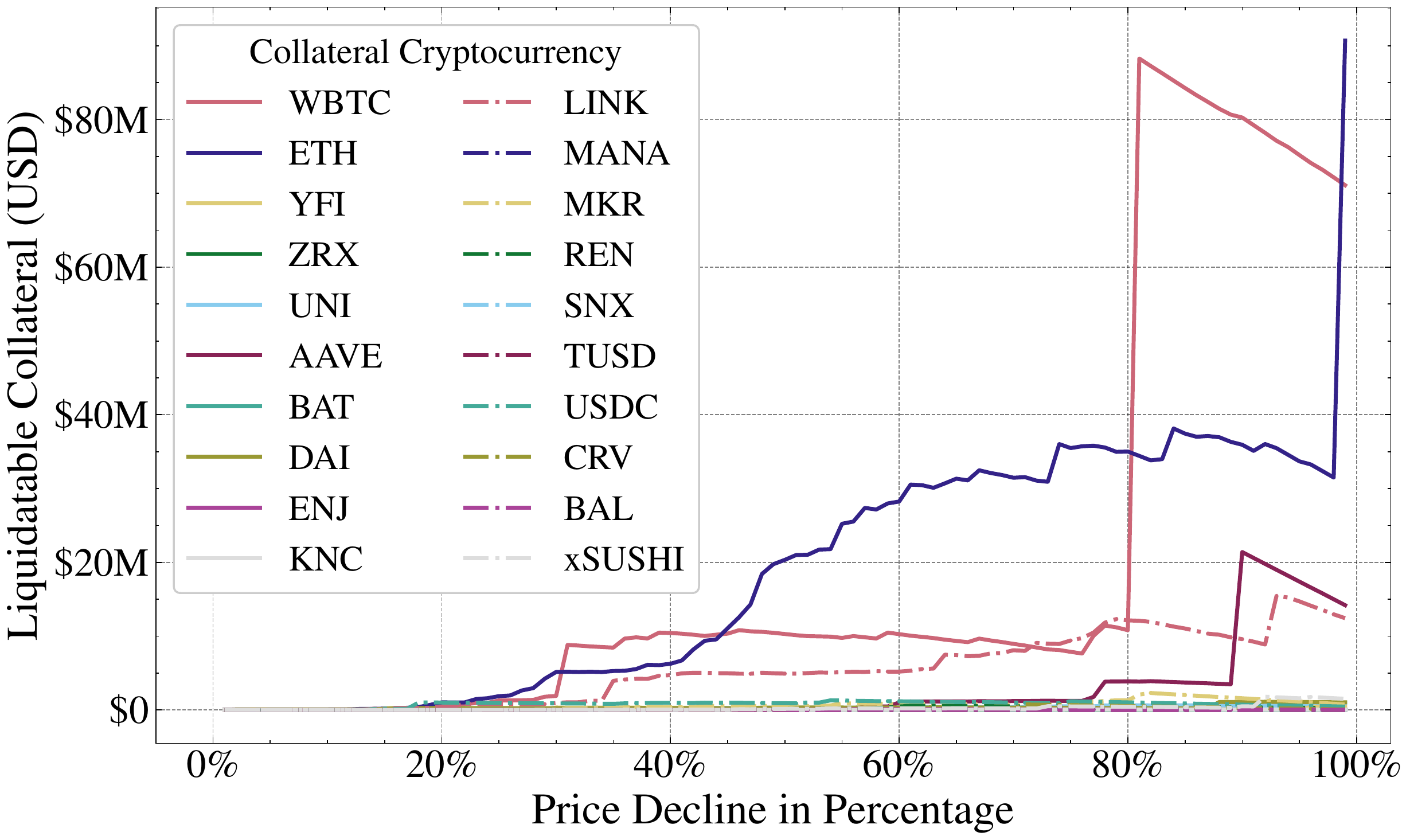}
         \caption{Aave V2}
         \label{fig:aave-liquidation-sensitivity}
     \end{subfigure}
     \begin{subfigure}[b]{0.495\textwidth}
         \centering
         \includegraphics[width=\textwidth]{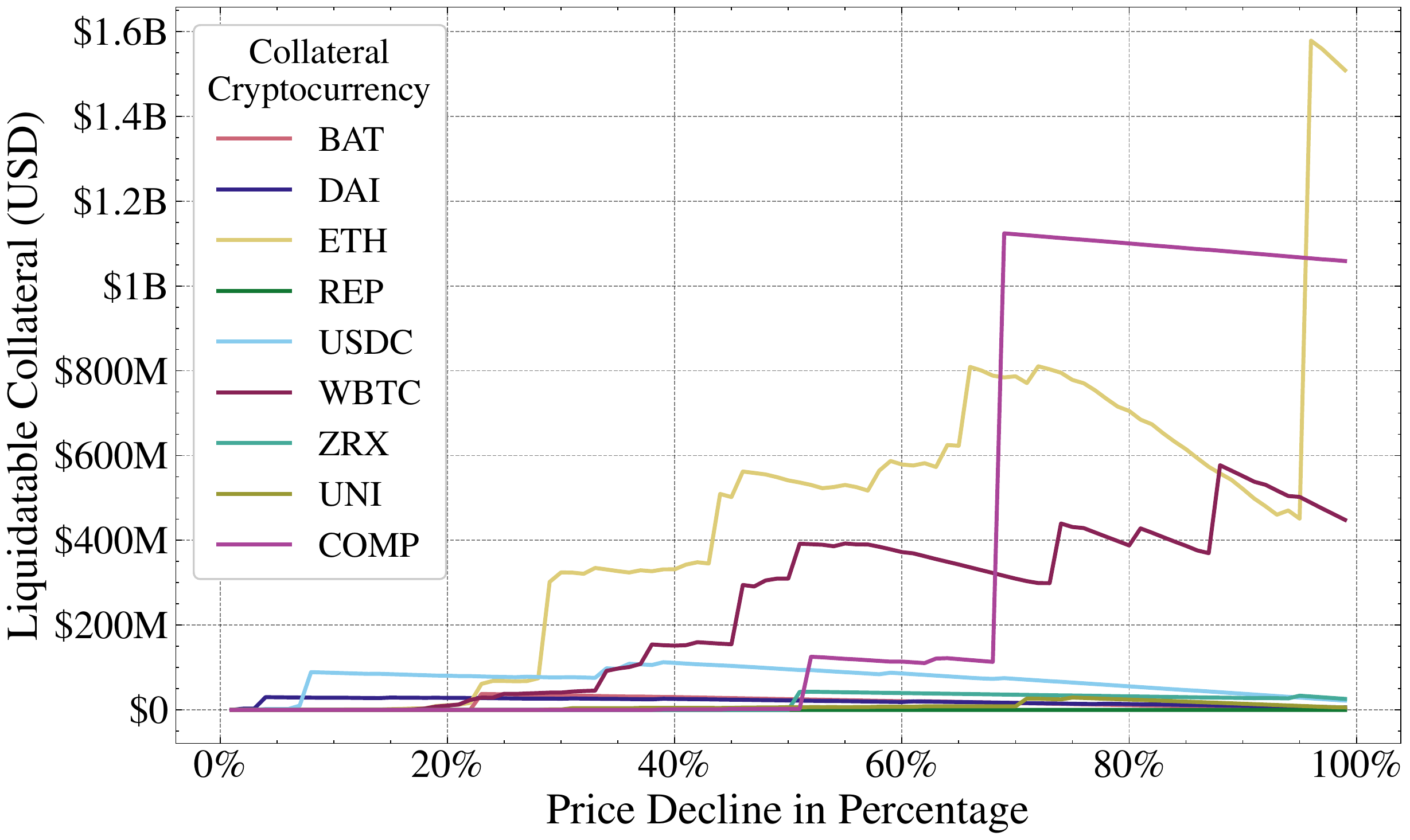}
         \caption{Compound}
         \label{fig:compound-liquidation-sensitivity}
     \end{subfigure}
     \begin{subfigure}[b]{0.495\textwidth}
         \centering
         \includegraphics[width=\textwidth]{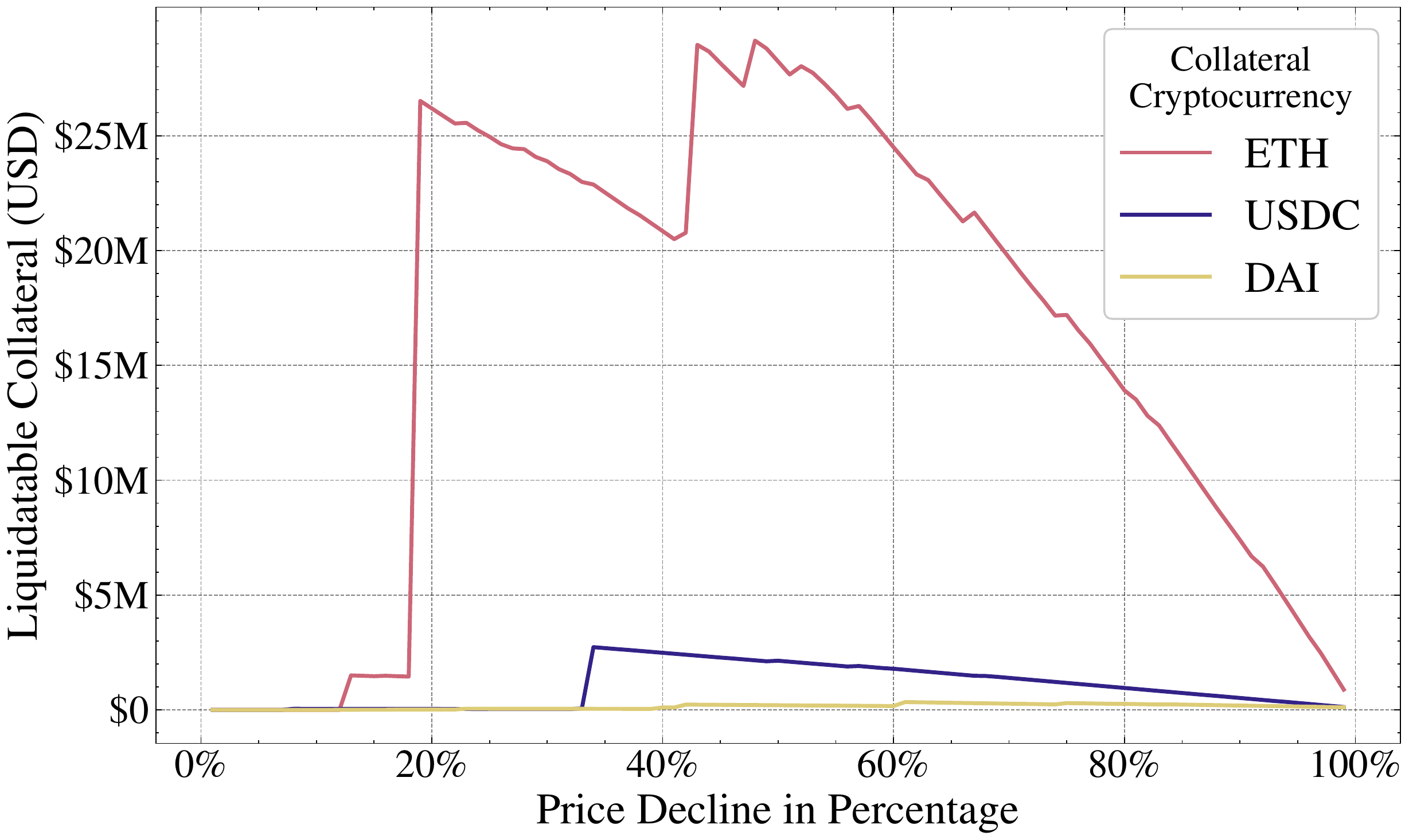}
         \caption{dYdX}
         \label{fig:dydx-liquidation-sensitivity}
     \end{subfigure}
     \begin{subfigure}[b]{0.495\textwidth}
         \centering
         \includegraphics[width=\textwidth]{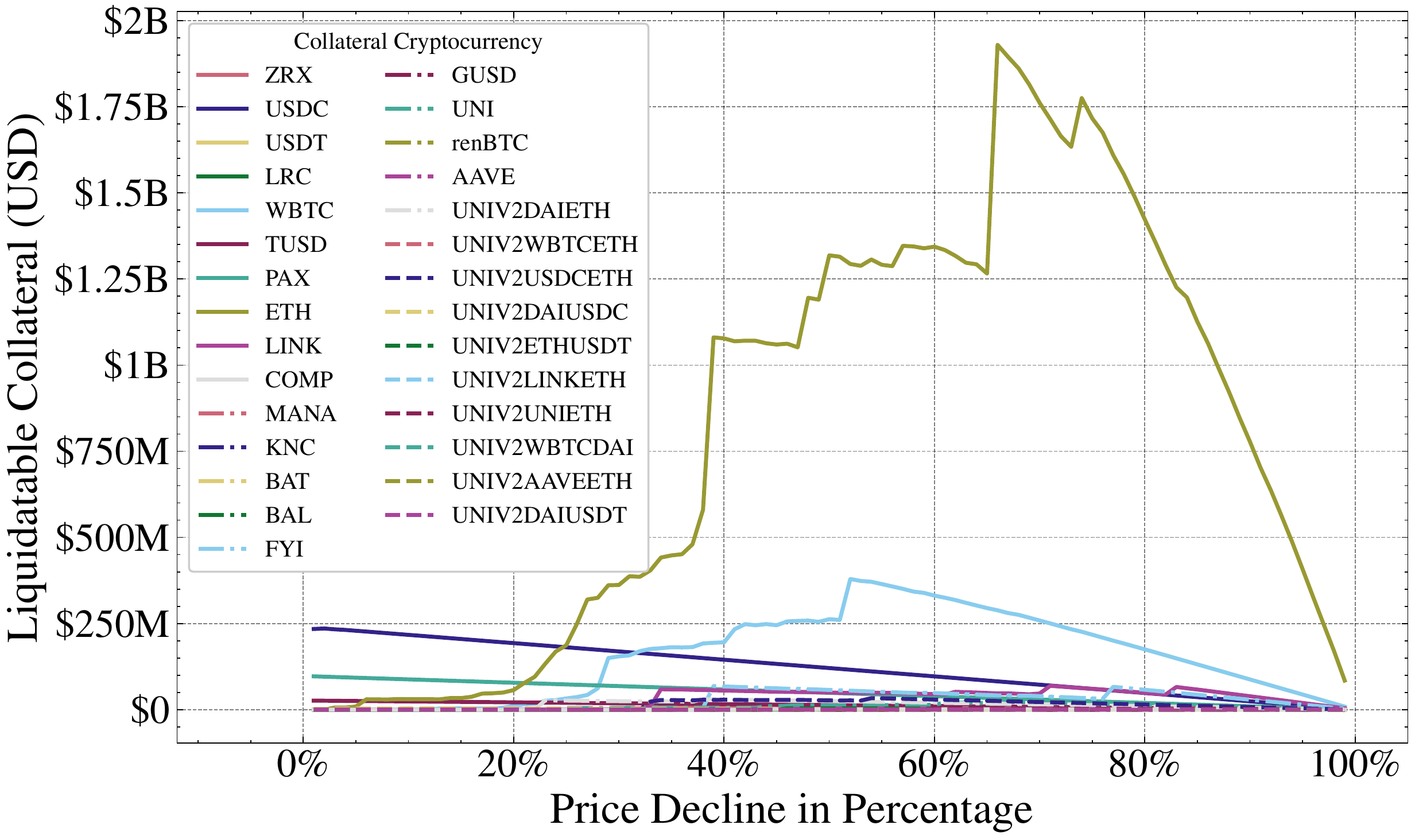}
         \caption{MakerDAO}
         \label{fig:maker-liquidation-sensitivity}
     \end{subfigure}
     \caption{Liquidation sensitivity to price decline in four lending platforms. Liquidation sensitivity denotes the amount of collateral that would be liquidated, if the price of the collateral would decline by up to $100$\%. We find that all of the four lending platforms are sensitive to the price decline of ETH. Although Aave V2 and Compound follow similar liquidation mechanisms and have similar TVL, Aave V2 is more stable to price declines.}
     \label{fig:liquidation-sensitivity}
\end{figure*}

We detail how we measure the liquidation sensitivity in Algorithm~\ref{alg:liquidationsensitivity}. Specifically, we examine whether each debt becomes liquidatable because of the price decline of the given cryptocurrency. Note that when counting the liquidatable collateral value, we consider the value decrease due to the price decline. We then present the sensitivity results in Figure~\ref{fig:liquidation-sensitivity}. We find that all of the four lending platforms are sensitive to the price decline of ETH. For example, an immediate $43$\% decline of the ETH price (analogous to the ETH price decline on the~13th of March,~2020), would result in up to~$1.07$B~USD collateral to become liquidatable on MakerDAO. To our surprise, although Aave V2 and Compound follow similar liquidation mechanisms and have similar TVL, Aave V2 is more stable to price declines in terms of liquidatable collateral. By manually inspecting, we find that this is because Aave V2 users prefer adopting a multiple-cryptocurreny collateral. Hence, the positions in Aave V2 are less likely to become liquidatable due to the price decline of a single cryptocurrency.

\subsubsection{Stability of Stablecoins}\label{sec:stabilityofstablecoins}
We observe that certain borrowers collateralize one stablecoin and borrow another stablecoin. Through such strategy, a borrower reduces the likelihood of liquidations, because the prices of stablecoins are deemed stable (cf.\ Section~\ref{sec:stablecoinbackground}). To measure the stability of this stablecoin borrowing strategy, we collect the prices of three popular stablecoins, DAI, USDC, and USDT, reported by the price oracle Chainlink~\cite{arijuel2017chainlink}, from block~\block{9976964} (May-01-2020) to~\block{12344944} (Apr-30-2021). We find that the price differences among the three stablecoins is within~$5\%$ in~$99.97\%$ of the measured ~$\numprint{2367981}$ blocks ($1$~year). This indicates that the aforementioned stablecoin borrowing strategy could have been stable for most of the time in 2020. However, we remark that liquidation risks still remain. The maximum price difference we detect is $11.1\%$ between USDC and DAI at block~\block{10578280}.

\subsection{Remarks}
Our measurements and analysis provide the following insights on DeFi liquidation mechanisms.
\begin{enumerate}
    \item Existing liquidation mechanisms generate remarkable financial rewards for liquidators (cf.\ Section~\ref{sec:profitandloss}). Liquidators are well incentivized to actively perform liquidations. This is confirmed by the severe gas price competition among liquidation transactions (cf.\ Section~\ref{sec:fixedspreadparticipation}) and short bid intervals in auction liquidations (cf.\ Section~\ref{sec:auctionparticipation}). However, the fixed spread liquidation allows liquidators to over-liquidate a borrowing position, which incurs unnecessary losses to the borrowers (cf.\ Section~\ref{sec:overliquidation}).
    \item Excessive transaction fees necessarily lead to unprofitable liquidation opportunities and Type II bad debt. Overdue liquidations increase the likelihood of Type I bad debt (cf.\ Section~\ref{sec:bad_debts} and Section~\ref{sec:unprofitable_liquidation}).
    \item Fixed spread liquidators can use flash loans to eliminate the risk of holding a specific asset (cf.\ Section~\ref{sec:flash_loan_usages}). Auction liquidators are exposed to the risk of price fluctuations during auctions, and may hence suffer a loss (cf.\ Appendix~\ref{app:postliquidationprice}).
    \item We show that at the time of writing, the studied lending platforms in this work (Aave V2, Compound, dYdX, and MakerDAO) are sensitive (i.e., the amount of the liquidatable collateral due to the price decline of a cryptocurrency) to the price decline of ETH (cf.\ Section~\ref{sec:sensitivity}).
    \item We evaluate the lending and borrowing practice when focussing exclusively on stablecoins. We show that such strategy mitigates the risk of liquidations for most of the time, while liquidations can still occur (cf.\ Section~\ref{sec:stabilityofstablecoins}).
\end{enumerate}

\section{Towards Better Liquidation Mechanisms}\label{sec:better-liquidation}
In this section, we objectively compare the studied liquidation mechanisms and present an optimal fixed spread liquidation strategy that aggravates the loss of borrowers.

\subsection{Objectively Comparing Liquidation Mechanisms}
Defining the optimality of a liquidation mechanism is challenging, because a liquidation process is a zero-sum game: the liquidator wins, what the borrower loses. As such, we can rather quantitatively reason about which liquidation mechanism is likely advantageous to a liquidator, or to a borrower.

\begin{figure*}[tb!]
    \centering
    \includegraphics[width=\textwidth]{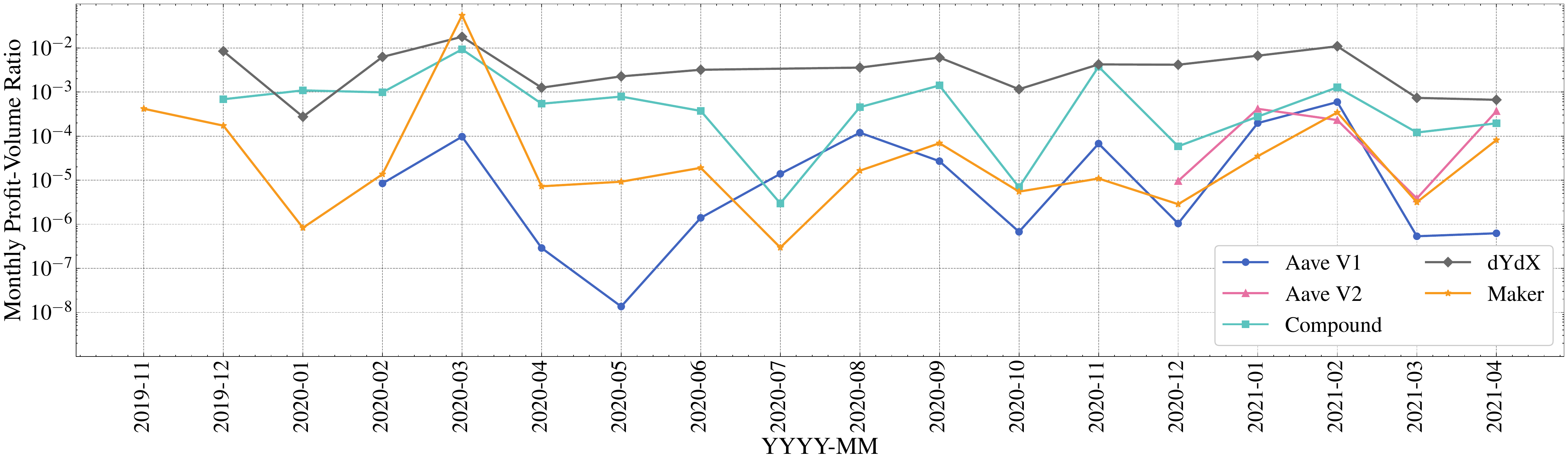}
    \caption{Comparison of the monthly liquidation profit over the monthly average collateral volume for the DAI/ETH lending markets. The lower the profit-volume ratio is, the better the liquidation protocol is for borrower.}
    \label{fig:profit_volume_ratio}
\end{figure*}

Based on our empirical data, we proceed by comparing the aforementioned liquidation mechanisms. We define the monthly \emph{profit-volume ratio} as the ratio between the monthly accumulated liquidation profit and the monthly average collateral volume. To avoid that our results are biased by the asset price fluctuations of different cryptocurrencies, we only study the liquidations repaid in DAI and collateralized in ETH, which are available across all the studied lending platforms. We present the monthly profit-volume ratios of the four platforms from November,~2019~to April,~2021~in Figure~\ref{fig:profit_volume_ratio}.

Our results show that dYdX has a higher profit-volume ratio than the other four platforms, meaning that dYdX is in expectation favorable to liquidators and worse for borrowers. This observation matches the fact that dYdX does not set a close factor, which means that a debt can be fully liquidated once its health factor declines below $1$. We further observe that MakerDAO consistently shows a smaller ratio than Compound, except for the outlier in March,~2020~(cf.\ Section~\ref{sec:insights}). Surprisingly, while Aave follows the same liquidation mechanism of Compound, the profit-volume ratio of Aave, especially Aave V1, remains below Compound. We infer that this is because the number of DAI/ETH liquidations events on Aave are rare (cf.\ Table~\ref{tab:DAI-ETH-liquidations}, Appendix~\ref{app:daiethliquidations})~–-~we hence believe that the Aave market is not sufficiently indicative to draw a representative conclusion. Overall our results suggest that the auction mechanism favors the borrowers more than a fixed spread liquidation with a close factor beyond $50\%$. 

\subsection{Optimal Fixed Spread Liquidation Strategy}\label{sec:optimalstrategy}
The configuration of a close factor (cf.\ Section~\ref{sec:terminology}) restricts the profit of the liquidator (i.e., the loss of the borrower) in a single fixed spread liquidation. We denote the strategy of liquidating up to the close factor limit within a single liquidation as the \emph{up-to-close-factor} strategy. Intuitively, a liquidator is rational to perform the up-to-close-factor strategy, because the profit is positively correlated to the liquidation amount. However, we find that a liquidator can lift the restriction of the close factor by performing two successive liquidations. The optimal strategy leverages the rule that a position remains liquidatable as long as it is in an unhealthy state, no matter this position has been liquidated or not previously.

In this optimal strategy, instead of pushing to close factor limit, the liquidator liquidates as much as possible but still keeps the position in an unhealthy state in the first liquidation. Then, in the second liquidation, the liquidator liquidates the remaining collateral up to the close factor. We detail the optimal fixed spread liquidation strategy in Algorithm~\ref{alg:optimalfixedspreadliquidation}. 

\begin{algorithm}[tb]
\SetAlgoLined
\SetKwProg{Fn}{Function}{:}{end}
\SetKwInOut{Input}{Input}
\SetKwInOut{Output}{Output}
\SetKwFunction{Liquidate}{Liquidate}
\SetKwFunction{IsLiquidatable}{Liquidatable}

\Input{A liquidatable position $\mathcal{POS}=\left \langle C, D \right \rangle$, where $C$ represents the collateral value, while $D$ represents the debt value; Liquidation threshold $\mathbf{LT}$; Liquidation spread $\mathbf{LS}$; Close factor $\mathbf{CF}$.}
\Output{Amount of debt to repay in the two optimal successive liquidations, $repay_1$ and $repay_2$.}
\BlankLine
\Fn{\IsLiquidatable{$\mathcal{POS}$}}{
    \KwRet{$\frac{\mathcal{POS}.C \times \mathbf{LT}}{\mathcal{POS}.D} > 1$}\;
}
\BlankLine
\Fn{\Liquidate{$\mathcal{POS}$, $repay$}}{
    $\mathcal{POS}'\leftarrow\left \langle C - repay\times (1+\mathbf{LS}) , D - repay \right \rangle$\;
    \KwRet{$\mathcal{POS}'$}\;
}
\BlankLine
$repay_1 \leftarrow \operatorname{argmax}_r$ \IsLiquidatable{\Liquidate{$\mathcal{POS}, r$}}\;
$\mathcal{POS}'\leftarrow$ \Liquidate{$\mathcal{POS}, repay_1$}\;
$repay_2 \leftarrow \mathcal{POS}'.D \times \mathbf{CF}$\;

\caption{Optimal fixed spread liquidation strategy.}
\label{alg:optimalfixedspreadliquidation}
\end{algorithm}

\subsubsection{Optimality Analysis} Given a liquidatable borrowing position with $C$ collateral value and $D$ debt (cf.\ Equation~\ref{eq:position}), we proceed to analyze the profit of our optimal strategy.

\begin{equation}\label{eq:position}
    \mathcal{POS} = \left \langle C, D \right \rangle
\end{equation}
$\mathbf{LT}$, $\mathbf{LS}$, $\mathbf{CR}$ denote the liquidation threshold, liquidation spread and close factor respectively (cf.\ Section~\ref{sec:terminology}).

Following Algorithm~\ref{alg:optimalfixedspreadliquidation}, the repaid debt amounts in the two successive liquidations are given in Equation~\ref{eq:repay1} and~\ref{eq:repay2}. Note that the $D-\mathbf{LT}\cdot C > 0$ because $\mathcal{POS}$ is liquidatable (i.e., the debt is greater than the borrowing capacity). We show in Appendix~\ref{app:reasonalconfiguration} that a reasonable fixed spread liquidation configuration satisfies $1-\mathbf{LT}(1+\mathbf{LS})>0$.

\begin{equation}\label{eq:repay1}
\begin{aligned}
    repay_1 &= \operatorname{argmax}_r \frac{\mathbf{LT}(C-r(1+\mathbf{LS}))}{D-r}\geq 1\\
    &=\frac{D-\mathbf{LT}\cdot C}{1-\mathbf{LT}(1+\mathbf{LS})}
\end{aligned}
\end{equation}

\begin{equation}\label{eq:repay2}
    repay_2 = \mathbf{CF}\left(D-repay_1\right)= \mathbf{CF}\left(D-\frac{D-\mathbf{LT}\cdot C}{1-\mathbf{LT}(1+\mathbf{LS})}\right)
\end{equation}
The overall profit of the two liquidations is shown in Equation~\ref{eq:overallprofit}.
\begin{equation}\label{eq:overallprofit}
\begin{aligned}
    profit_{o} &= (repay_1 + repay_2)\times \mathbf{LS}\\
    &=\mathbf{LS}\cdot\mathbf{CF}\cdot D+\mathbf{LS}(1-\mathbf{CF})\left(\frac{D-\mathbf{LT}\cdot C}{1-\mathbf{LT}(1+\mathbf{LS})}\right)
\end{aligned}
\end{equation}

If the liquidator instead chooses to perform the up-to-close-factor strategy, the repay amount is $\mathbf{CF}\cdot D$ and the profit hence is $profit_{c}=\mathbf{LS}\cdot\mathbf{CF}\cdot D$. Therefore, the optimal strategy can yield more profit than the up-to-close-factor strategy. The increase rate of the liquidation profit is shown in Equation~\ref{eq:increaserate}.

\begin{equation}\label{eq:increaserate}
    \Delta R_{profit} = \frac{profit_o-profit_c}{profit_c}=\frac{\mathbf{CF}}{1-\mathbf{CF}}\cdot\frac{1-\mathbf{LT}\cdot\mathsf{CR}}{1-\mathbf{LT}(1+\mathbf{LS})}
\end{equation}
where $\mathsf{CR}=\frac{C}{D}$ is the collateralization ratio (cf.\ Section~\ref{sec:terminology}). We notice that the optimal strategy is more effective when $\mathsf{CR}$ is low. 

\subsubsection{Case Study}
In the following, we study the most profitable fixed spread liquidation transaction (\BiggestSingleLiquidationProfit)\footnote{Transaction hash: \etherscantx{0x53e09adb77d1e3ea593c933a85bd4472371e03da12e3fec853b5bc7fac50f3e4}} we detect and showcase how the optimal fixed spread liquidation strategy increases the profit of a liquidator. In this Compound liquidation, the liquidator first performs an oracle price update\footnote{Compound allows any entity to update the price oracle with authenticated messages signed by, for example, off-chain price sources.}, which renders a borrowing position liquidatable. The liquidator then liquidates the position within the same transaction. In Table~\ref{tab:casestudystatus}, we present the status change of the position following the price update. Before the price update (block~\block{11333036}), the position owns a total collateral of $135.07$M~USD (with a borrowing capacity of $101.30$M~USD), and owes a debt of~$101.18$M~USD. After the price of DAI increases from $1.08$ to $1.095299$~USD/DAI, the total debt reaches $102.61$M~USD, while the borrowing capacity is only $102.55$M~USD. The health factor drops below~$1$, and hence the position becomes liquidatable.

\begin{table}[tb!]
\centering
\caption{Status of the borrowing position (\account{0x909b443761bbD7fbB876Ecde71a37E1433f6af6f}). Note we ignore the tiny amount of collateral and debt in USDT that the borrower owns and owes. The liquidation thresholds (i.e., $\mathbf{LT}$) of DAI and USDC are both~$0.75$.}
\resizebox{\columnwidth}{!}{%
\begin{tabular}{@{}ccc|c|c@{}}
\toprule
\multirow{2}{*}{\bf Token} & \multirow{2}{*}{\bf Collateral} & \multirow{2}{*}{\bf Debt} & \multicolumn{2}{c}{\bf Price (USD)} \\ \cmidrule(l){4-5} 
&  &  & Block \block{11333036} & After Price Update \\ \midrule
DAI & $108.51$M & $93.22$M & $1.08$ & $1.095299$ \\
USDC & $17.88$M & $506.64$k & $1$ & $1$ \\ \midrule\midrule
\multicolumn{3}{c|}{\bf Total Collateral (USD)} & $135.07$M & $136.73$M \\
\multicolumn{3}{c|}{\bf Borrowing Capacity (USD)} & $101.30$M & $102.55$M \\
\multicolumn{3}{c|}{\bf Total Debt (USD)} & $101.18$M & $102.61$M \\ \bottomrule
\end{tabular}%
}
\label{tab:casestudystatus}
\end{table}

\begin{table}[tb!]
\centering
\caption{The depiction of liquidation strategies. At the time of the original liquidation, the price of DAI is $1.095299$~USD/DAI. The close factor is $50\%$. The optimal strategy is the most profitable liquidation mechanism for the liquidator.}
\resizebox{\columnwidth}{!}{%
\begin{tabular}{@{}c|l|l@{}}
\toprule
\textbf{Original liquidation} & \multicolumn{2}{l}{\begin{tabular}[c]{@{}l@{}}Repay $46.14$M~USD\\ Receive $49.83$M~DAI\\ \emph{Profit} $3.69$M~DAI\end{tabular}} \\ \midrule\midrule
\textbf{Up-to-close-factor strategy} & \multicolumn{2}{l}{\begin{tabular}[c]{@{}l@{}}Repay $46.61$M~DAI\\ Receive $50.34$M~DAI\\ \emph{Profit} $3.73$M~DAI\end{tabular}} \\ \midrule\midrule
\multirow{4}{*}{\textbf{Optimal strategy}} & Liquidation 1 & Liquidation 2 \\ \cmidrule(l){2-3} 
 & \begin{tabular}[c]{@{}l@{}}Repay $296.61$K~DAI\\ Receive $320.34$K~DAI\\ \emph{Profit} $23.73$K~DAI\end{tabular} & \begin{tabular}[c]{@{}l@{}}Repay $46.46$M~DAI\\ Receive $50.18$M~DAI\\ \emph{Profit} $3.72$M~DAI\end{tabular} \\ \bottomrule
\end{tabular}%
}
\label{tab:liquidation-strategy-comparisons}
\end{table}

To evaluate the up-to-close-factor strategy and our optimal liquidation strategy, we implement the original liquidation and the two liquidation strategies\footnote{We publish the smart contract code at \url{https://anonymous.4open.science/r/An-Empirical-Study-of-DeFi-Liquidations-Anonymous/CompoundLiquidationCaseStudy.sol}.} in Solidity v0.8.4\footnote{\url{https://docs.soliditylang.org/en/v0.8.4/}}.
We execute them on the corresponding blockchain states\footnote{We fork the Ethereum mainchain locally from block~\block{11333036} and apply all the transactions executed prior to the original liquidation transaction in block~\block{11333037}. We then execute the liquidation strategies to ensure that they are validated on the exact same state of the original liquidation.} and present the results in Table~\ref{tab:liquidation-strategy-comparisons}. We find that the optimal strategy is superior to the up-to-close-factor strategy and can generate an additional profit of $49.26$K~DAI (\OptimalLiquidationAdditionalProfit) compared to the original liquidation.

\subsubsection{Mitigation}
The aforementioned optimal strategy defeats the original intention of a close factor, which incurs undesirably additional losses to borrowers. A possible mitigation solution is that for every position only one liquidation is permitted within one block. Such a setting enforces a liquidator adopting the optimal strategy to settle the two liquidations in two blocks, which decreases the success probability.

We proceed to assume the existence of a mining liquidator with a mining power $\alpha$. Given a liquidation opportunity, the up-to-close-factor strategy produces a profit of $profit_c$, while the optimal strategy yields $profit_{o_1}$ and $profit_{o_2}$ respectively in the two successive liquidations. We further assume that there is no ongoing consensus layer attack (e.g., double-spending), implying that a miner with an $\alpha$ fraction mining power mines the next block is with a probability of $\alpha$. We hence derive the the expected profit of the two strategies as shown in Equation~\ref{eq:expected-profit-close} and~\ref{eq:expected-profit-optimal}\footnote{To ease understanding, we assume other competing liquidators adopt the up-to-close-factor strategy.}.

\begin{equation}\label{eq:expected-profit-close}
    \mathbb{E}[\text{up-to-close-factor}] = \alpha\cdot profit_c
\end{equation}
\begin{equation}\label{eq:expected-profit-optimal}
    \mathbb{E}[\text{optimal}] = \alpha\cdot profit_{o_1} + \alpha^2\cdot profit_{o_2}
\end{equation}
The liquidator is incentivized to perform the optimal strategy only when $\mathbb{E}[\text{optimal}] > \mathbb{E}[\text{up-to-close-factor}]$, leading to Equation~\ref{eq:optimal-rational-condition}.

\begin{equation}\label{eq:optimal-rational-condition}
\alpha > \frac{profit_c-profit_{o_1}}{profit_{o_2}}
\end{equation}
Intuitively, $profit_{o_1}$ is relatively small compared to $profit_{c}$ and $profit_{o_2}$ because the liquidator needs to keep the position unhealthy after the first liquidation. The expected profit in the second liquidation then should be sufficient to cover the opportunity cost in the first one, which is typically unattainable.
Instantiating with our case study liquidation (cf.\ Table~\ref{tab:casestudystatus}), we show that a rational mining liquidator would attempt the optimal strategy in two consecutive blocks only if its mining power is over $99.68\%$. Therefore, we conclude that the one liquidation in one block effectively reduces the expected profit of the optimal liquidation strategy, protecting borrowers from a further liquidation losses.

\section{Related Work}\label{sec:related-work}

\point{Blockchains and DeFi}
There is a growing body of literature on blockchains and DeFi. Qin \etal~\cite{qin2020attacking} study flash loan attacks and present an optimization approach to maximize the profit of DeFi attacks. Zhou \etal~\cite{zhou2020high} analyze sandwich attacks in decentralized exchanges. Eskandari \etal~\cite{eskandari2019sok} provide an overview of the blockchain front-running attacks. Daian \etal~\cite{daian2019flash} investigate the front-running attacks in decentralized exchanges and propose the concept of \emph{Miner Extractable Value} (MEV), a financial revenue miners can extract through transaction order manipulation. Qin \etal~\cite{qin2021quantifying} quantify the extracted MEV on the Ethereum blockchain, including fixed spread liquidations, and present a generalized front-running algorithm, transaction replay. Zhou \etal~\cite{zhou2021just} propose a framework called \textsc{DeFiPoser} that allows to automatically create profit-generating transactions given the blockchain state.

\point{Blockchain Borrowing and Lending Markets}
Darlin \etal~\cite{darlin2020optimal} study the MakerDAO liquidation auctions. The authors optimize the costs for participating in the auctions and find that most auctions conclude at higher than optimal prices. The work appears real-world relevant, as it considers the transaction fees, conversion costs and cost of capital, yet it does not consider potential gas bidding contests by the end of MakerDAO auctions~\cite{daian2019flash}. Kao \etal~\cite{kao2020analysis} and ZenGo~\cite{zengo-compound} are to our knowledge the first to have investigated Compound's liquidation mechanism (the third biggest lending protocol in terms of USD at the time of writing). Perez~\etal~\cite{perez2020liquidations} follow up with a report that focuses on additional on-chain analytics of the Compound protocol. DragonFly Research provides a blog post~\cite{medium-liquidation} about the liquidator profits on Compound, dYdX and MakerDAO. Minimizing financial deposit amounts in cryptoeconomic protocols, while maintaining the same level of security is studied in Balance~\cite{harzbalance}.

\point{Liquidations in Traditional Finance} Liquidations are essential to traditional finance (TradFi) and are well studied in the related literature~\cite{titman1984effect,shleifer1992liquidation,alderson1995liquidation,almgren1999value,reinhart2011liquidation}. We remark that liquidations in blockchain systems are fundamentally different from those in TradFi in terms of high-level designs and settlement mechanisms. 


\section{Conclusion}\label{sec:conclusion}
Due to their significant volatility when compared to alternative financial vehicles cryptocurrencies are attracting speculators. Furthermore, because speculators seek to further their risk exposure, non-custodial lending and borrowing protocols on blockchains are thriving. The risks of borrowing, however, manifests themselves in the form of liquidation profits claimed by liquidators.

In this paper we study the lending platforms that capture $85$\% of the blockchain lending market. We systematize the most prevalent liquidation mechanisms and find that many liquidations sell excessive amounts of borrower's collateral. In this work we provide extensive data analytics covering over $2$ years the prevalent $4$ lending protocols. We systematize their respective liquidation mechanisms and show that most liquidation systems are unfavorable to the borrowers. We finally show an optimal liquidation strategy which we have not yet observed in the wild.




\bibliographystyle{ACM-Reference-Format}
\bibliography{references}


\appendix

\section{Post-Liquidation Price Movement Measurement}\label{app:postliquidationprice}
Auction liquidators are exposed to the risk that the price of collateral further declines during the auction, which may cause a loss to the liquidator. In the following, we study the price movement of collateral with respect to the debt currency after the settlement of a fixed spread liquidation and after the initiation of an auction liquidation. For each observed liquidation, we record the block-by-block oracle prices for a duration of~$\numprint{1440}$ blocks, which corresponds to about~$6$ hours. We summarize the following patterns of the post-liquidation price movement.

\begin{itemize}
\item \textbf{Horizontal:} The collateral price does not change after liquidation (e.g., the oracle price is not updated).

\item \textbf{Rise:} Within $\numprint{1440}$ blocks, the collateral price remains higher than the liquidation price.

\item \textbf{Fall:} Within $\numprint{1440}$ blocks, the collateral price remains lower than the liquidation price.

\item \textbf{Rise-Fall} Within $\numprint{1440}$ blocks, the collateral price first rises beyond the liquidation price, then falls below the liquidation price.

\item \textbf{Fall-Rise} Within $\numprint{1440}$ blocks, the collateral price declines below the liquidation price, and then rises above the liquidation price.

\item \textbf{Rise-Fluctuation:} Within $\numprint{1440}$ blocks, the collateral price first rises beyond, then declines below the liquidation price, and repeats such movement more than twice.

\item \textbf{Fall-Fluctuation:} Within $\numprint{1440}$ blocks, the collateral price first declines below, then rises beyond the liquidation price, and repeats such movement more than twice.
\end{itemize}

We observe that the collateral price remains below the liquidation price by the end of the observation window for only $19.07\%$ of the \TotalSuccessfulLiquidations observed liquidations. If those liquidations would have been conducted through an auction, the liquidator might have suffered a loss.



\begin{table}[bt!]
    \centering
    \caption{Observed collateral price movements, measured in relation to the liquidation price.}
    \resizebox{1\columnwidth}{!}{%
    \begin{tabular}{lccc}
    \toprule
    \bf Price Movement    &  \bf Liquidations  & \bf Maximum Price  &  \bf Minimum Price\\
    \midrule
      Horizontal &   \HorizontalLiquidations & - & - \\
      Rise  & \RiseLiquidations  &  \RisePeak  & - \\
      Fall  &  \FallLiquidations  & - & \FallValley  \\
      Rise-Fall  & \RiseFallLiquidations  &  \RiseFallPeak  &  \RiseFallValley \\
      Fall-Rise &  \FallRiseLiquidations  & \FallRisePeak &  \FallRiseValley \\
      Rise-Fluctuation & \RiseFluctuationLiquidations  & \RiseFluctuationPeak  & \RiseFluctuationValley \\
      Fall-Fluctuation  &  \FallFluctuationLiquidations & \FallFluctuationPeak  & \FallFluctuationValley \\
    \bottomrule
    \end{tabular}%
    }
    
    \label{tab:postliquidation-price-movement}
\end{table}

\section{Monthly DAI/ETH Liquidations}\label{app:daiethliquidations}
In Table~\ref{tab:DAI-ETH-liquidations}, we show the number of monthly liquidations that are repaid in DAI and collateralized in ETH on Aave V1, Aave V2, Compound, dYdX, and MakerDAO.
\begin{table}[htb!]
\centering
\caption{Number of monthly liquidations for the DAI/ETH lending markets to compare the five platforms on Figure~\ref{fig:profit_volume_ratio}.}
\resizebox{\columnwidth}{!}{%
\begin{tabular}{lrrrrr}
\toprule
\multicolumn{1}{c}{\multirow{2}{*}{\textbf{Year-Month}}} & \multicolumn{5}{c}{\textbf{Number of Liquidations}} \\ \cmidrule{2-6} 
\multicolumn{1}{c}{} & Aave V1 & Aave V2 & Compound & dYdX & MakerDAO \\ \midrule
2019-11 & 0     & 0     & 0     & 0     & 119 \\
2019-12 & 0     & 0     & 29    & 203   & 118 \\
2020-01 & 0     & 0     & 21    & 25    & 12 \\
2020-02 & 7     & 0     & 63    & 467   & 105 \\
2020-03 & 31    & 0     & 712   & 1124  & 4222 \\
2020-04 & 1     & 0     & 27    & 57    & 9 \\
2020-05 & 7     & 0     & 29    & 241   & 24 \\
2020-06 & 7     & 0     & 24    & 127   & 45 \\
2020-07 & 6     & 0     & 22    & 0     & 20 \\
2020-08 & 9     & 0     & 37    & 98    & 42 \\
2020-09 & 25    & 0     & 99    & 192   & 105 \\
2020-10 & 2     & 0     & 16    & 39    & 11 \\
2020-11 & 8     & 0     & 95    & 144   & 31 \\
2020-12 & 6     & 1     & 23    & 226   & 20 \\
2021-01 & 22    & 20    & 76    & 570   & 62 \\
2021-02 & 13    & 41    & 108   & 334   & 246 \\
2021-03 & 2     & 10    & 15    & 58    & 11 \\
2021-04 & 5     & 16    & 33    & 28    & 212 \\
\bottomrule
\end{tabular}%
}
\label{tab:DAI-ETH-liquidations}
\end{table}

\section{Reasonable Fixed Spread Liquidation Configurations}\label{app:reasonalconfiguration}
In the following, we analyze reasonable configurations of the liquidation threshold $\mathbf{LT}$ and the liquidation spread $\mathbf{LS}$ (cf.\ Section~\ref{sec:background}). We assume a borrowing position $\mathcal{POS} = \left \langle C, D \right \rangle$, where $C$ is the value of the collateral while $D$ is the value of the debt. We consider the situation that the health factor of $\mathcal{POS}$ drops below $1$ (cf.\ Equation~\ref{eq:healthfactorbelow1}), i.e., $\mathcal{POS}$ becomes liquidatable.
\begin{equation}\label{eq:healthfactorbelow1}
    \mathsf{HF} = \frac{C\cdot\mathbf{LT}}{D} < 1
\end{equation}

A liquidator then liquidates $\mathcal{POS}$ by repaying a debt of $r$. In return the liquidator receives a collateral of $r\cdot(1+\mathbf{LS})$. After the liquidation, the health factor becomes Equation~\ref{eq:newhealthfactor}.
\begin{equation}\label{eq:newhealthfactor}
    \mathsf{HF}' = \frac{(C-r\cdot(1+\mathbf{LS}))\mathbf{LT}}{D-r}
\end{equation}

Ideally, a liquidation should help to increase the health factor of a borrowing position (cf.\ Equation~\ref{eq:increasehealthfactor}).
\begin{equation}\label{eq:increasehealthfactor}
    \mathsf{HF}'>\mathsf{HF}
\end{equation}

Following Equation~\ref{eq:healthfactorbelow1},~\ref{eq:newhealthfactor}, and~\ref{eq:increasehealthfactor}, we obtain Equation~\ref{eq:healthfactorincreasecondition}.
\begin{equation}\label{eq:healthfactorincreasecondition}
    1 + \mathbf{LS} < \frac{C}{D}
\end{equation}
Note that $\mathbf{LS}$ is positive. Therefore, when $\mathcal{POS}$ is under-collateralized (i.e., $\frac{C}{D}<1$), Equation~\ref{eq:healthfactorincreasecondition} can never be satisfied. This implies that a fixed spread liquidation never increase the health factor of a under-collateralized position.

In the case that $\mathcal{POS}$ is over-collateralized (i.e., $\frac{C}{D}>1$) but liquidatable (cf.\ Equation~\ref{eq:healthfactorbelow1}), if we configure $\mathbf{LT}$ and $\mathbf{LS}$ satisfying that $\mathbf{LT}(1+\mathbf{LS}) \geq 1$,
this configuration then conflicts with Equation~\ref{eq:healthfactorincreasecondition} because following Equation~\ref{eq:healthfactorincreasecondition} we obtain Equation~\ref{eq:conflicts}.
\begin{equation}\label{eq:conflicts}
    \mathbf{LT}(1 + \mathbf{LS})< \frac{C\cdot\mathbf{LT}}{D} = \mathsf{HF} < 1
\end{equation}

We therefore conclude that $1-\mathbf{LT}(1+\mathbf{LS}) > 0$ is the prerequisite such that a fixed spread liquidation can increase the health factor of an over-collateralized liquidatable borrowing position.

\end{document}